\documentclass[12pt]{iopart}

\usepackage{iopams}
\usepackage{amssymb}
\usepackage{textcomp}
\usepackage{graphicx}

\begin{document}

\title[Dielectric response of electron-hole systems]{Dielectric response of electron-hole systems. Nondegenerate case and quantum corrections}

\author{D Semkat$^1$, H Stolz$^2$, W-D Kraeft$^2$ and H Fehske$^1$}

\address{$^1$ Institut f\"ur Physik, Ernst-Moritz-Arndt-Universit\"at Greifswald, Felix-Hausdorff-Str.\ 6, 17489 Greifswald, Germany}
\address{$^2$ Institut f\"ur Physik, Universit\"at Rostock, Albert-Einstein-Str.\ 23-24, 18059 Rostock, Germany}
\ead{dirk.semkat@uni-greifswald.de}
%\vspace{10pt}
%\begin{indented}
%\item[]August 2017
%\end{indented}

\begin{abstract}
Analytical results for the dielectric function in RPA are derived for three-, two-, and one-dimensional semiconductors in the weakly-degenerate limit. Based on this limit, quantum corrections are derived. Further attention is devoted to systems with linear carrier dispersion and the resulting Dirac-cone physics.
\end{abstract}

%\noindent{\it Keywords}: dielectric response, electron-hole plasma
%\maketitle

\section{Introduction}

The response of a medium to external electromagnetic fields is determined by its dielectric function \cite{mahan}. The description of various phenomena is closely connected with this quantity, e.g. the buildup of screening or the behavior of bound states of electrons and holes in a semiconductor (excitons) surrounded by a plasma of free carriers \cite{gruenesbuch,kremp2005}.
While in earlier times the main interest was focused on higher densities (see, e.g., \cite{arndt96}), with the observation of Rydberg excitons in cuprous oxide (Cu$_2$O) \cite{nature2014,heckoetter2018} the behavior in a very low density plasma has become important, as Rydberg states due to their large Bohr radius (up to 1 micrometer for $n=30$ \cite{versteegh2021}), are extremely sensitive to such a low density plasma. In a recent study we have shown that the Mott effect, i.e., the vanishing of a bound state at a certain plasma density, occurs for quantum number $n=25$ already at a density of $10^{8}$ cm$^{-3}$.   Most important, however, is that, despite the low density, the commonly used Debye approximation \cite{gruenesbuch} gives completely wrong results \cite{semkat2019,semkat2021}. Crucial in these calculations was the use of an exact expression for the dielectric function.  To extend these studies to two-dimensional systems such as transition metal dichalcogenide monolayers \cite{raja2017} and even to one-dimensional systems, one has to know the exact dielectric function also for lower dimensionality.

A further example are the collective oscillation modes of this plasma (plasmons), the complex energies of which are given by the zeros of the complex dielectric function. 
Knowledge on this function is, therefore, essential for the understanding of various optical and transport properties of semiconductors.

In the current paper we derive and discuss analytical results for the dielectric function in RPA (\textit{random phase approximation}). Section 3 is devoted to the case of bulk systems. Afterwards in section 4 we discuss lower-dimensional systems. Finally, in section 5 we show an extension of the results to moderate degeneracy, i.e., derive quantum corrections to the nondegenerate limit.

Some derivations are presented in the appendices, e.g., the effective Coulomb potential for quasi-two-dimensional systems (appendix A) and the dielectric function for the one-dimensional case (B).
The quite distinct case of linear quasiparticle dispersion, which occurs in graphene~\cite{CGP09} and topological quantum matter~\cite{HK10} with Dirac-cone functionality is analysed on the same footing in appendix C.

\section{Polarization function}

The dielectric function of an electron-hole plasma in a $d$-dimensional semiconductor is connected via
\begin{equation}\label{epsilon}
\varepsilon(k,\omega)=1-\sum\limits_{a=\mathrm{e,h}}V_{aa}(k)\,\Pi_{aa}(k,\omega)
\end{equation}
with the polarization function $\Pi_{aa}$  of electrons and holes ($a=$ e,h), respectively \cite{gruenesbuch}. $V_{aa}$ is the dimension-dependent interaction potential between carriers of species $a$.
In RPA, the polarization function is given by the well-known Lindhard expression \cite{gruenesbuch}
\begin{eqnarray}\label{Pi}
\Pi_{aa}(k,\omega)&=&(2s_a+1)\int\frac{\mathrm{d}^dq}{(2\pi)^d}
\frac{f_a\left(\frac{\mathbf{k}}{2}-\mathbf{q}\right)-f_a\left(\frac{\mathbf{k}}{2}+\mathbf{q}\right)}
{E_a\left(\frac{\mathbf{k}}{2}-\mathbf{q}\right)-E_a\left(\frac{\mathbf{k}}{2}+\mathbf{q}\right)+\hbar\omega+\mathrm{i}\epsilon}\,.
\end{eqnarray}
Here $f_a$ is the distribution function of species $a$ which is, in the nondegenerate case, given by the Boltzmann distribution
\begin{equation}\label{fboltzmann}
f_a(k)=\frac{n_a\Lambda_a^d}{2s_a+1}\,\exp\left(-\frac{\hbar^2k^2}{2m_ak_{\rm B}T_a}\right)
\end{equation}
with temperature $T_a$ and density $n_a$ of species $a$. $\Lambda_a$ is the thermal deBroglie wavelength,
\begin{equation}
\Lambda_a=\left(\frac{2\pi\hbar^2}{m_ak_{\rm B}T_a}\right)^{1/2}
\end{equation}
and $2s_a+1$ is the spin degeneracy factor which is for electrons and holes $2s_a+1=2$.
The quasiparticle energies $E_a$ are approximated by free particle energies \cite{rigidshift} which read, in the usual approximation of parabolic bands,
\begin{equation}\label{pardisp}
E_a(k)=\frac{\hbar^2k^2}{2m_a}+\mathrm{Re}\Sigma_a(\mathbf{k},\omega)\Big|_{\omega=E_a(\mathbf{k})/\hbar}\approx\frac{\hbar^2k^2}{2m_a}\,,
\end{equation}
It is well known and frequently cited that the integral in (\ref{Pi}) can be evaluated in the nondegenerate case (for $d=3$) analytically \cite{fehr94,seidel95,arndt96}. Reference \cite{klimontovich74} is regarded as the key source, however, the explicit derivation cannot be found in that work. Moreover, in the mentioned papers, the three-dimensional (3d) case is (implicitly) assumed. 
We will, therefore, rederive the ``classical'' result in 3d and consider afterwards the 2d and 1d cases.

\section{Bulk semiconductor}

In the case of a bulk semiconductor, the interaction potential between carriers of species $a$ $V_{aa}$ $(a=\mathrm{e,h})$ is given by the three-dimensional Coulomb potential $V_{aa}^{\rm C}$,
\begin{equation}
V_{aa}^{\rm C}(k)=\frac{e_ae_a}{\epsilon_0\epsilon_b}\frac{1}{k^2}=\frac{e^2}{\epsilon_0\epsilon_b}\frac{1}{k^2}
\end{equation}
with $\epsilon_b$ being the background dielectric constant.
The integral in (\ref{Pi}) can be written in spherical coordinates $(q,\vartheta,\varphi)$. We lay the $z$-axis of the $\mathbf{q}$-integration into $\mathbf{k}$. The difference in the numerator of (\ref{Pi}) then becomes
\begin{eqnarray}\label{fdiff}
f_a\left(\frac{\mathbf{k}}{2}-\mathbf{q}\right)-f_a\left(\frac{\mathbf{k}}{2}+\mathbf{q}\right)
=\frac{n_a\Lambda_a^3}{2}\,\exp\left[-\frac{\hbar^2}{2m_ak_{\rm B}T_a}\left(\frac{k^2}{4}+q^2\right)\right]
\nonumber\\
\times
\left[\exp\left(\frac{\hbar^2}{2m_ak_{\rm B}T_a}kq\,\cos\,\vartheta\right)-\exp\left(-\frac{\hbar^2}{2m_ak_{\rm B}T_a}kq\,\cos\,\vartheta\right)\right]\!.
\end{eqnarray}
Using (\ref{pardisp}) the energy difference in the denominator of (\ref{Pi}) is
\begin{eqnarray}\label{Ediff}
E_a\left(\frac{\mathbf{k}}{2}-\mathbf{q}\right)-E_a\left(\frac{\mathbf{k}}{2}+\mathbf{q}\right)
&=\frac{\hbar^2}{2m_a}\left[\left(\frac{\mathbf{k}}{2}-\mathbf{q}\right)^2-\left(\frac{\mathbf{k}}{2}+\mathbf{q}\right)^2\right]
\nonumber\\
&=-\frac{\hbar^2}{m_a}kq\,\cos\,\vartheta\,.
\end{eqnarray}
Inserting the differences (\ref{fdiff}) and (\ref{Ediff}) into (\ref{Pi}), substituting in the usual manner $\cos\,\vartheta=t$ and performing the trivial $\varphi$-integration, the polarization function reads
\begin{eqnarray}\label{Pi1}
\fl\Pi_{aa}^{(3d)}(k,\omega)=\frac{1}{(2\pi)^2}n_a\Lambda_a^3\,\exp\left(-\frac{\hbar^2k^2}{8m_ak_{\rm B}T_a}\right)
\nonumber\\
\fl\times
\int\limits_0^{\infty}\mathrm{d}q\,q^2\,\exp\left(-\frac{\hbar^2q^2}{2m_ak_{\rm B}T_a}\right)
\int\limits_{-1}^1\mathrm{d}t\,\frac{\exp\left(\frac{\hbar^2}{2m_ak_{\rm B}T_a}kqt\right)-\exp\left(-\frac{\hbar^2}{2m_ak_{\rm B}T_a}kqt\right)}
{\hbar\omega-\frac{\hbar^2}{m_a}kqt+\mathrm{i}\epsilon}\,.
\end{eqnarray}
For the following calculations, we introduce the abbreviations $\beta=1/(k_{\rm B}T_a)$, $a=\hbar^2/(2m_a)$, and $w=\hbar\omega$ and substitute $akq=x$. The double integral to be evaluated reads now
\begin{eqnarray}\label{I1}
I=\frac{1}{a^3k^3}\int\limits_0^{\infty}\mathrm{d}x\,x^2\,\exp\left(-\frac{\beta}{ak^2} x^2\right)
\int\limits_{-1}^1\mathrm{d}t\,\frac{\mathrm{e}^{\beta xt}-\mathrm{e}^{-\beta xt}}
{w-2xt+\mathrm{i}\epsilon}\,.
\end{eqnarray}
We separate real and imaginary parts by expanding with the complex conjugate of the denominator,
\begin{eqnarray}\label{I1a}
I&=&\frac{1}{a^3k^3}\int\limits_0^{\infty}\mathrm{d}x\,x^2\,\exp\left(-\frac{\beta}{ak^2} x^2\right)
\int\limits_{-1}^1\mathrm{d}t\,\left(\mathrm{e}^{\beta xt}-\mathrm{e}^{-\beta xt}\right)\nonumber\\
&&\times\left\{
\frac{w-2xt}{(w-2xt)^2+\epsilon^2}
-\frac{\mathrm{i}\,\epsilon}
{(w-2xt)^2+\epsilon^2}\right\}\nonumber\\
&=&I_1+I_2
\,.
\end{eqnarray}

The imaginary part $I_2$ can be evaluated easily. We perform in this term the limit $\epsilon\to 0$, obtaining
\begin{eqnarray}\label{I1aa}
I_2&=&-\frac{\mathrm{i}\pi}{a^3k^3}\int\limits_0^{\infty}\mathrm{d}x\,x^2\,\exp\left(-\frac{\beta}{ak^2} x^2\right)
\int\limits_{-1}^1\mathrm{d}t\,\left(\mathrm{e}^{\beta xt}-\mathrm{e}^{-\beta xt}\right)
\delta(w-2xt)\,.
\end{eqnarray}
Performing subsequently $t$- and $x$-integration yields
\begin{eqnarray}\label{I1b}
I_2=-\frac{\mathrm{i}\pi}{4\beta a^2k}\left(\mathrm{e}^{\beta w/2}-\mathrm{e}^{-\beta w/2}\right)\exp\left(-\frac{\beta w^2}{4a k^2}\right)
\,.
\end{eqnarray}

In its present form, the real part $I_1$ cannot be solved straightforwardly. Therefore, we introduce an auxiliary integral by making use of
\begin{equation}\label{trick}
\frac{1}{y}=\int\limits_0^{\infty}\mathrm{d}s\,\mathrm{e}^{-ys}\qquad\mbox{for}\;y>0\,.
\end{equation}
After rearranging the exponentials, changing the order of integrations, and substituting $4s=z$, integral $I_1$ then reads
\begin{eqnarray}\label{I3}
I_1=&\frac{1}{4a^3k^3}\int\limits_0^{\infty}\mathrm{d}z\,\exp\left[-\frac{1}{4}(w^2+\epsilon^2)z\right]\int\limits_0^{\infty}\mathrm{d}x\,
x^2\,\exp\left(-\frac{\beta}{ak^2} x^2\right)
\nonumber\\
&\times
\int\limits_{-1}^1\mathrm{d}t\,\exp\left(-x^2zt^2\right)
\left\{\exp\left[(wz+\beta) xt\right]-\exp\left[(wz-\beta) xt\right]\right\}\nonumber\\
&\times(w-2xt)\,.
\end{eqnarray}
The $t$-, $x$-, and $z$-integrals can now be performed subsequently yielding (note that the limit $\epsilon\to 0$ can be done trivially)
\begin{eqnarray}\label{I6}
\fl I_1=-\frac{\sqrt{\pi}}{2\beta a^2k}\,\exp\left(\frac{\beta ak^2}{4}\right)
\left\{
F\left[\frac{\sqrt{\beta}}{2\sqrt{a}k}\left(w+ak^2\right)\right]
-F\left[\frac{\sqrt{\beta}}{2\sqrt{a}k}\left(w-ak^2\right)\right]
\right\},
\end{eqnarray}
where $F$ denotes Dawson's integral which is closely connected with the confluent hypergeometric function $_1F_1$ (also referred to as Kummer function) and with the Faddeeva function (or Kramp function) $\mathrm{w}$ \cite{abramowitz},
\begin{eqnarray}\label{dawsonkummer}
F(x)=\int\limits_0^x\mathrm{d}t\,\exp\left(t^2-x^2\right)=x\,_1F_1\left(1,\frac{3}{2};-x^2\right)=\frac{\sqrt{\pi}}{2}\,\mathrm{Im}\,\mathrm{w}(x)\,.
\end{eqnarray}
The latter function is in turn connected to the complementary complex error function,
\begin{eqnarray}\label{faddeeva}
\mathrm{w}(x)=\exp\left(-x^2\right)\,\mathrm{erfc}(-\mathrm{i}x)\,,
\end{eqnarray}
i.e., its real part is given by $\mathrm{Re}\,\mathrm{w}(x)=\exp\left(-x^2\right)$ $(x\in\mathbb{R})$. Therefore, we can combine (\ref{I6}) and (\ref{I1b}),
\begin{eqnarray}\label{Ifinal}
\mathrm{e}^{-\frac{\beta ak^2}{4}}(I_1+I_2)
=\nonumber\\
=-\frac{\pi}{4\beta a^2k}
\Bigg(
\frac{2}{\sqrt{\pi}}\left\{
F\left[\frac{\sqrt{\beta}}{2\sqrt{a}k}\left(w+ak^2\right)\right]
-F\left[\frac{\sqrt{\beta}}{2\sqrt{a}k}\left(w-ak^2\right)\right]
\right\}\nonumber\\
+\mathrm{i}\left\{
\mathrm{exp}\left[-\frac{\beta}{4ak^2}\left(w-ak^2\right)^2\right]
-\mathrm{exp}\left[-\frac{\beta}{4ak^2}\left(w+ak^2\right)^2\right]
\right\}
\Bigg)\nonumber\\
=\frac{\mathrm{i}\pi}{4\beta a^2k}
\Bigg\{
\mathrm{w}\left[\frac{\sqrt{\beta}}{2\sqrt{a}k}\left(w+ak^2\right)\right]
-\mathrm{w}\left[\frac{\sqrt{\beta}}{2\sqrt{a}k}\left(w-ak^2\right)\right]
\Bigg\}\,.
\end{eqnarray}

Inserting the result (\ref{Ifinal}) into the polarization function (\ref{Pi1}) we get for the latter quantity
\begin{eqnarray}\label{Pi2}
\Pi_{aa}^{(3d)}(k,\omega)
=\nonumber\\
=\frac{1}{(2\pi)^2}n_a\Lambda_a^3\,\frac{\mathrm{i}\pi}{4\beta a^2k}
\Bigg\{
\mathrm{w}\left[\frac{\sqrt{\beta}}{2\sqrt{a}k}\left(w+ak^2\right)\right]
-\mathrm{w}\left[\frac{\sqrt{\beta}}{2\sqrt{a}k}\left(w-ak^2\right)\right]
\Bigg\}\nonumber\\
=\mathrm{i}\frac{\sqrt{\pi}}{2}n_a\sqrt{\frac{2m_a}{\hbar^2k^2}}\frac{1}{\sqrt{k_{\rm B}T_a}}
\Bigg\{
\mathrm{w}\left[\frac{1}{2\sqrt{k_{\rm B}T_a}}\sqrt{\frac{2m_a}{\hbar^2k^2}}\left(\hbar\omega+\frac{\hbar^2k^2}{2m_a}\right)\right]\nonumber\\
-\mathrm{w}\left[\frac{1}{2\sqrt{k_{\rm B}T_a}}\sqrt{\frac{2m_a}{\hbar^2k^2}}\left(\hbar\omega-\frac{\hbar^2k^2}{2m_a}\right)\right]
\Bigg\}\,.
%\nonumber\\
\end{eqnarray}
We finally use the dimensionless quantities
\begin{eqnarray}
x=\frac{1}{2}\frac{\omega/\omega_{\rm e}}{k/\kappa}\,,\quad
y=\left(\frac{\hbar^2k^2}{8m_{\rm e}k_{\rm B}T_a}\right)^{1/2}\,,\quad
s=\left(\frac{m_{\rm h}}{m_{\rm e}}\right)^{1/2}\,,\quad
\kappa=\left(\frac{2n_{\rm e}e^2}{\epsilon_0\epsilon_bk_{\rm B}T_a}\right)^{1/2}\,,\nonumber\\
\omega_{\rm e}=\left(\frac{n_{\rm e}e^2}{\epsilon_0\epsilon_bm_{\rm e}}\right)^{1/2}\,,
\end{eqnarray}
where $\kappa$ and $\omega_{\rm e}$ are inverse screening length and plasma frequency of the electrons, respectively. Summing up $\Pi_{\rm ee}$ and $\Pi_{\rm hh}$, one arrives at
\begin{eqnarray}\label{epsilon-3d}
\varepsilon(k,\omega)=&1-\mathrm{i}\frac{\sqrt{\pi}}{4}\frac{\kappa^2}{k^2}\nonumber\\
&\times\left[
\frac{\mathrm{w}(x+y)-\mathrm{w}(x-y)}{2y}+
\frac{\mathrm{w}(sx+y/s)-\mathrm{w}(sx-y/s)}{2y/s}
\right]\,.
\end{eqnarray}
For the first time, this result has been derived in \cite{klimontovich74}. In the form of (\ref{epsilon-3d}), it agrees with (19) in \cite{seidel95}.

\begin{figure}[h]%
\begin{center}
\includegraphics*[width=0.49\textwidth]{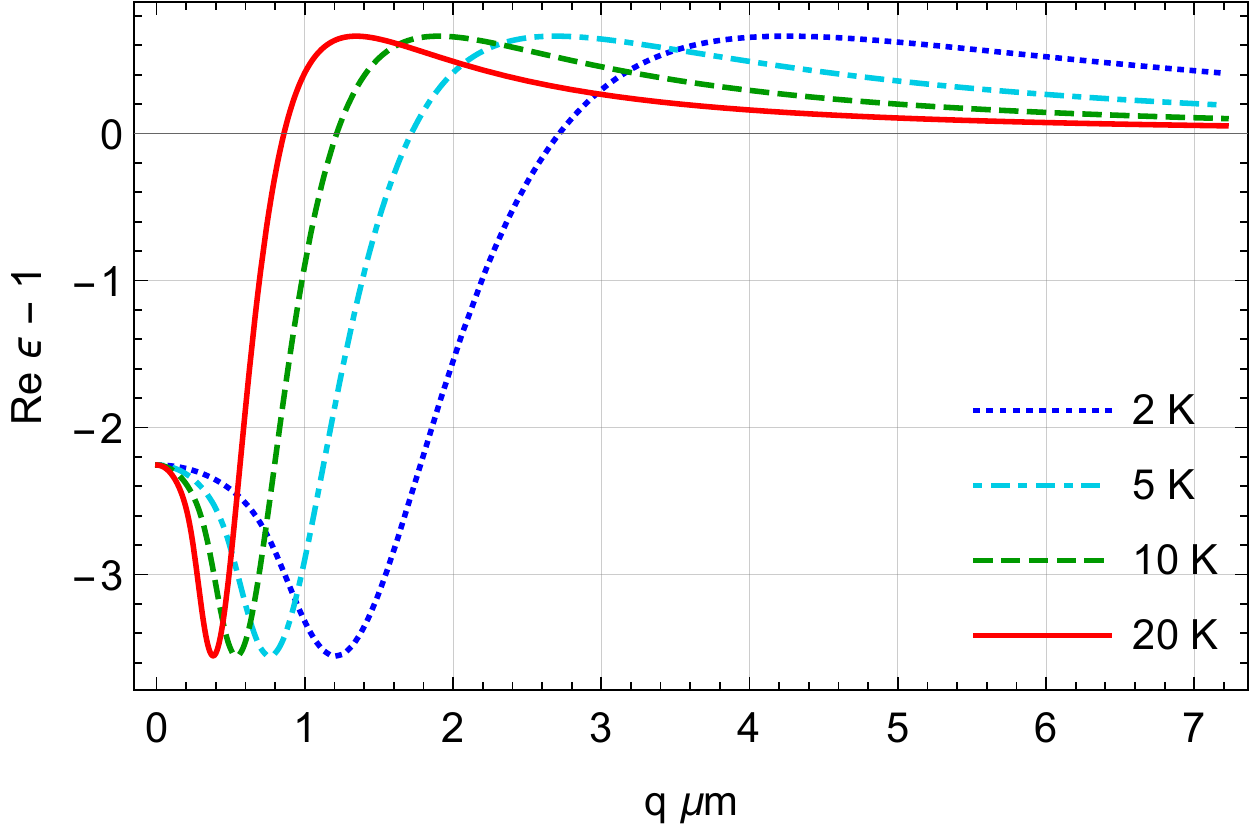}\hspace*{0.1cm}
\includegraphics*[width=0.49\textwidth]{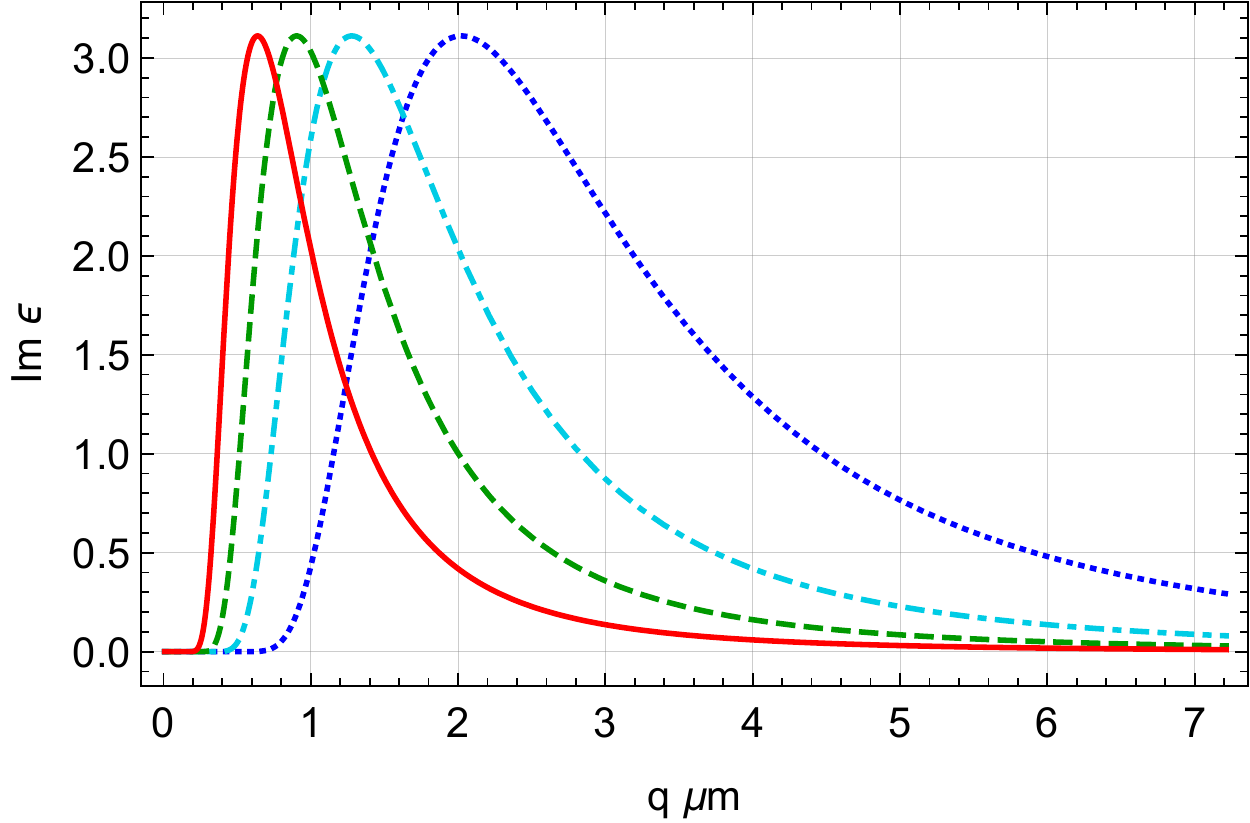}\\
\includegraphics*[width=0.49\textwidth]{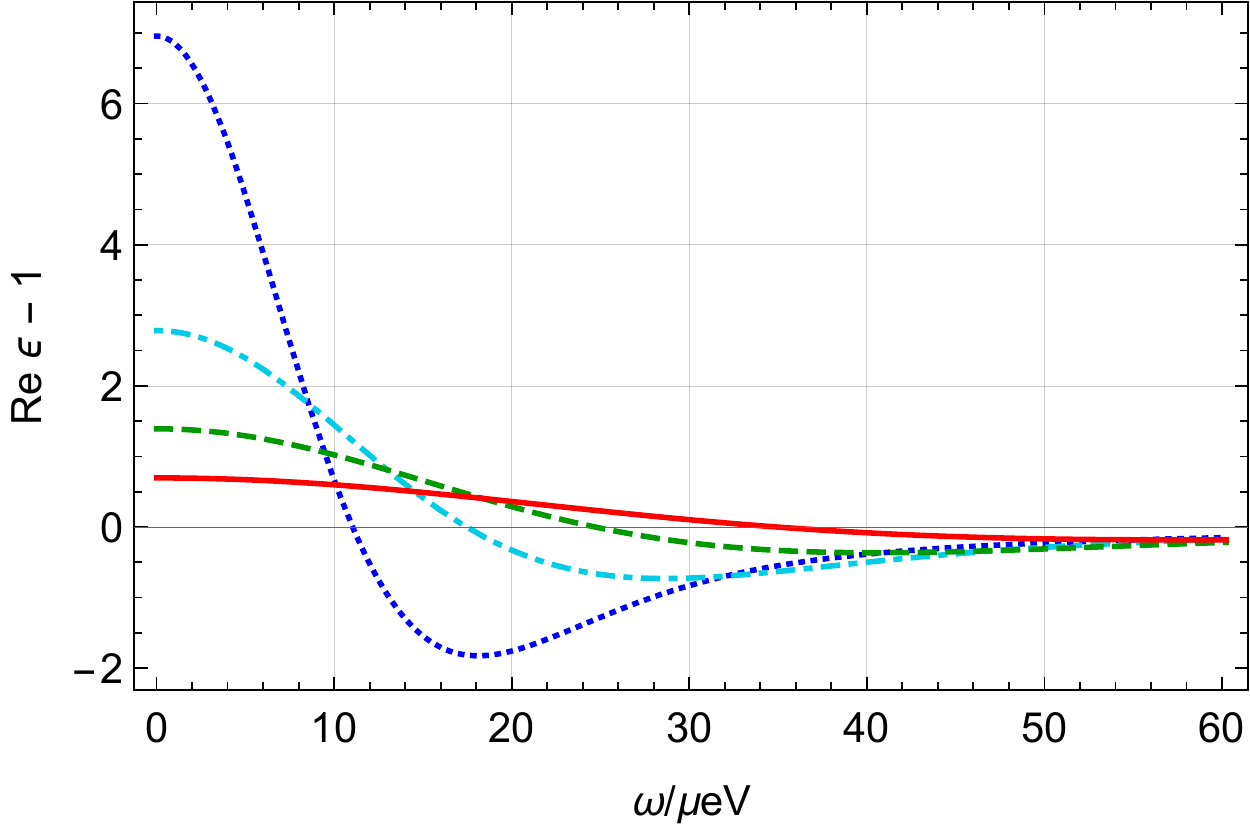}\hspace*{0.1cm}
\includegraphics*[width=0.49\textwidth]{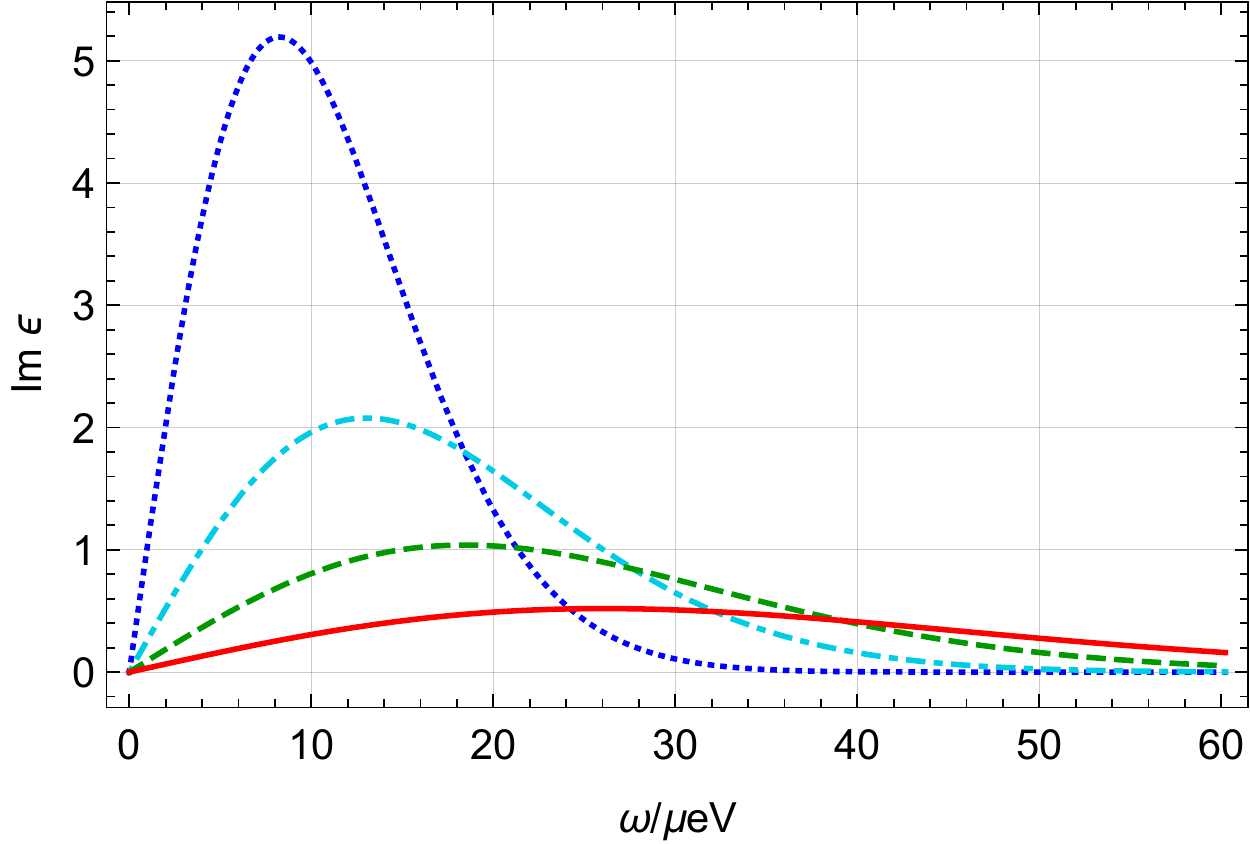}
\caption{Real (left panels) and imaginary parts (right panels) of the dielectric function \textit{vs.}\ wave number $q$ at a frequency of $\hbar\omega=15$ \textmu eV  (upper row) and \textit{vs.}\ frequency at a wave number of $q=2$ \textmu m$^{-1}$ (lower row) for electrons and holes in bulk cuprous oxide, each for a carrier density of $n=10^{12}$ cm$^{-3}$ and several temperatures.
}
\label{fig:epsilon-q-T}
\end{center}
\end{figure}
Figure \ref{fig:epsilon-q-T} shows real and imaginary parts of $\varepsilon(q,\omega)-1$. We consider electrons and holes in bulk Cu$_2$O with the parameters: electron mass $m_{\rm e}=0.985\,m_0$, hole mass $m_{\rm h}=0.575\,m_0$, and dielectric constant $\epsilon_{\rm b}=7.507$. Since figure \ref{fig:epsilon-q-T} serves here mainly for illustration of the considered quantity, we only briefly mention the contained physical information, i.e., the dispersion of collective plasma modes (plasmons) given by the zeros of Re $\varepsilon$ and their damping connected with Im $\varepsilon$.

\section{Two-dimensional semiconductor structures}

In two dimensions, the Coulomb potential is proportional to $\mathrm{log}\,r$ instead of $1/r$ as in 3d (corresponding to $1/k$ instead of $1/k^2$). Semiconductor structures like GaAs/AlGaAs quantum wells or TMDC monolayers are quasi-two-dimensional, i.e., the layer widths are small compared to their in-plane extension.

One possibility to handle this quasi-two-dimensionality is to use for the interaction potential $V_{aa}$ the Rytova--Keldysh potential \cite{rytova1967,keldysh1979}. Usually given in its quite complicated form in configuration space, it reads in momentum space simply \cite{selig2018}
\begin{equation}\label{vrk}
V_{aa}^{\rm RK}(q)=\frac{e^2}{2\epsilon_0\epsilon_{\rm sub}A}\,\frac{1}{q(1+r_0q)}\,,
\end{equation}
where $\epsilon_{\rm sub}$ is the mean dielectric constant of the substrates and $r_0$ the screening length, $r_0=d_0\epsilon_{\perp}/\epsilon_{\rm sub}$ with $\epsilon_{\perp}$ being the dielectric constant of the monolayer and $d_0$ its thickness. Depending on the latter parameter, the potential (\ref{vrk}) interpolates between the limiting cases of bulk material ($r_0q\gg 1$) and true 2d system ($r_0q\to 0$). 

Another way is to account for the confinement in the third dimension by the respective eigenfunctions of the carriers in the quantum well \cite{girndt1997,manzke2012},
\begin{equation}
V_{ab}(q)=\frac{e_ae_b}{2\epsilon_0\epsilon_{\rm b,w}q}\int\limits_{-\infty}^{\infty}\mathrm{d}z\int\limits_{-\infty}^{\infty}\mathrm{d}z'
\left|\phi_a(z)\right|^2\left|\phi_b(z')\right|^2\,\exp\left(-q|z-z'|\right),
\end{equation}
where $\phi_{a/b}$ are the wave functions of the motion in confinement- ($z$-)direction, $\epsilon_{\rm b,w}$ is the background dielectric constant in the well, and $e_{\rm e/h}=\mp e$.
An analytical expression for this effective quasi-two-dimensional potential is derived in \ref{veff2d}.

%\subsection{Parabolic carrier dispersion}

We consider a 2d system again with parabolic carrier dispersion. In that case, polar coordinates $(q,\varphi)$ seem to be convenient for the integral in (\ref{Pi}), however, Cartesian coordinates $(q_x,q_y)$ turn out to be the appropriate choice. The polarization function then reads
\begin{eqnarray}\label{Pi1-2d}
\Pi_{aa}^{(2d)}(k,\omega)=\frac{1}{2\pi}n_a\Lambda_a^2\,\exp\left(-\frac{\hbar^2k^2}{8m_ak_{\rm B}T_a}\right)\,I
\end{eqnarray}
with
\begin{eqnarray}\label{I1a-2d-cart-1}
I=&\frac{1}{a^2k_xk_y}
\int\limits_{-\infty}^{\infty}\mathrm{d}x\int\limits_{-\infty}^{\infty}\mathrm{d}y\,
\exp\left[-(ux^2+vy^2)\right]
\,\left[\mathrm{e}^{\beta(x+y)}-\mathrm{e}^{-\beta(x+y)}\right]
\nonumber\\
&\times\left\{
\frac{w-2(x+y)}{[w-2(x+y)]^2+\epsilon^2}-
\frac{\mathrm{i}\,\epsilon}{[w-2(x+y)]^2+\epsilon^2}\right\}\nonumber\\
=&I_1+I_2
\,,
\end{eqnarray}
where we have introduced the abbreviations $x=ak_xq_x,\;y=ak_yq_y,\;u=\beta/(ak_x^2),$ and $v=\beta/(ak_y^2)$.

Like in the previous section, we look at first at the second (imaginary) contribution $I_2$. Performing the limit $\epsilon\to 0$ and substituting $s=2(x+y)$ (i.e. $(x,y)\to(x,s)$) leads to
\begin{eqnarray}\label{I2-2d-cart}
I_2=&-\frac{\mathrm{i}\pi}{2a^2k_xk_y}\int\limits_{-\infty}^{\infty}\mathrm{d}x\int\limits_{-\infty}^{\infty}\mathrm{d}s\,
\exp\left\{-\left[ux^2+v\left(\frac{s}{2}-x\right)^2\right]\right\}
\nonumber\\
&\times
\left[\exp\left(\frac{\beta s}{2}\right)-\exp\left(-\frac{\beta s}{2}\right)\right]\delta(w-s)
\end{eqnarray}
which yields straightforwardly
\begin{eqnarray}\label{I2-2d-cart-1}
I_2=-\frac{\mathrm{i}\sqrt{\pi^3}}{2\sqrt{\beta a^3}k}\left[\exp\left(\frac{\beta w}{2}\right)-\exp\left(-\frac{\beta w}{2}\right)\right]
\,\exp\left(-\frac{\beta w^2}{4ak^2}\right)
\,.
\end{eqnarray}

In the first (real) contribution to $I$ (\ref{I1a-2d-cart-1}) we apply again the integration trick (\ref{trick}),
\begin{eqnarray}\label{I11-2d-cart}
I_1&=&\frac{1}{a^2k_xk_y}\int\limits_0^{\infty}\mathrm{d}z\,\exp\left[-(w^2+\epsilon^2)z\right]
\int\limits_{-\infty}^{\infty}\mathrm{d}x\int\limits_{-\infty}^{\infty}\mathrm{d}y\,\exp\left[-(u+4z)x^2\right]\nonumber\\
&&\times
\exp\left[-(v+4z)y^2\right]\exp\left(-8xyz\right)[w-2(x+y)]
\nonumber\\
&&\times\left\{\exp\left[(\beta+4wz)(x+y)\right]-\exp\left[-(\beta-4wz)(x+y)\right]\right\}\,.
\end{eqnarray}
Now the integrations can be performed subsequently leading to the result
\begin{eqnarray}\label{I11-2d-cart-3}
\fl I_1=\frac{\pi}{\sqrt{\beta a^3}k}\,\exp\left(\frac{\beta ak^2}{4}\right)
\left\{
F\left[\frac{\sqrt{\beta}}{2\sqrt{a}k}\left(w-ak^2\right)\right]
-F\left[\frac{\sqrt{\beta}}{2\sqrt{a}k}\left(w+ak^2\right)\right]
\right\}
\,,\nonumber
\end{eqnarray}
where $F$ again denotes Dawson's integral.

Analogously to (\ref{Ifinal}), we can sum up the real and imaginary parts and express them in terms of the Faddeeva function w so that we finally get for the polarization function
\begin{eqnarray}\label{Pi2-2d}
\Pi_{aa}^{(2d)}(k,\omega)
%=\nonumber\\
&=&\frac{1}{(2\pi)^2}n_a\Lambda_a^2\,\frac{\mathrm{i}\sqrt{\pi^3}}{2\sqrt{\beta a^3}k}\nonumber\\
&&\times\Bigg\{
\mathrm{w}\left[\frac{\sqrt{\beta}}{2\sqrt{a}k}\left(w+ak^2\right)\right]
-\mathrm{w}\left[\frac{\sqrt{\beta}}{2\sqrt{a}k}\left(w-ak^2\right)\right]
\Bigg\}\nonumber\\
&=&\mathrm{i}\frac{\sqrt{\pi}}{2}n_a\sqrt{\frac{2m_a}{\hbar^2k^2}}\frac{1}{\sqrt{k_{\rm B}T_a}}
\Bigg\{
\mathrm{w}\left[\frac{1}{2\sqrt{k_{\rm B}T_a}}\sqrt{\frac{2m_a}{\hbar^2k^2}}\left(\hbar\omega+\frac{\hbar^2k^2}{2m_a}\right)\right]\nonumber\\
&&-\mathrm{w}\left[\frac{1}{2\sqrt{k_{\rm B}T_a}}\sqrt{\frac{2m_a}{\hbar^2k^2}}\left(\hbar\omega-\frac{\hbar^2k^2}{2m_a}\right)\right]
\Bigg\}\,.
%\nonumber\\
\end{eqnarray}

Comparing the 3d and 2d results (\ref{Pi2}) and (\ref{Pi2-2d}), we see that both cases have the same form, but note the different character and dimension of $n_a$ -- bulk density \textit{vs.} area density), i.e., $\Pi_{aa}^{(2d)}(k,\omega)=\Pi_{aa}^{(3d)}(k,\omega)$.

A very similar, straightforward calculation in the 1d case yields the corresponding result (see Appendix B), i.e., the functional form of the RPA polarization function of the electron-hole plasma in the nondegenerate limit is independent on the dimensionality of the system.

%\subsection{Linear carrier dispersion}

In all cases considered above we assumed the usual parabolic approximation for valence and conduction bands leading to free-particle-like dispersions of electrons and holes. There are, however, quasi-two-dimensional systems (the probably most prominent being graphene) where the band structure gives rise to linear carrier dispersions ($E(\mathbf{k})=\gamma k$) and exhibits so-called Dirac cones near the charge neutrality point. The polarization function in such systems is usually considered in the highly degenerate limiting case \cite{shung1986,HD07,kotov2012,wunsch2006}, however, an analytical expression for a model system with linear dispersion in the case of weak degeneracy can be obtained, too, see Appendix C.

% One obtains for the imaginary part $\mathrm{Im}\,\Pi_{aa}$
% \begin{eqnarray}\label{ImPilin5}
% \mathrm{Im}\,\Pi_{aa}(k,\omega)
% =&-\frac{k}{16\pi\gamma}\,n_a\Lambda_a^2
% \left[\exp\left(\frac{\beta\hbar\omega}{2}\right)-\exp\left(-\frac{\beta\hbar\omega}{2}\right)\right]\\
% %\nonumber\\
% &\times
% \left\{
% \frac{2}{\beta\gamma k}
% \frac{K_1\left(\frac{\beta\gamma k}{2}\right)}{\sqrt{1-\left(\frac{\hbar\omega}{\gamma k}\right)^2}}
% +K_0\left(\frac{\beta\gamma k}{2}\right)\sqrt{1-\left(\frac{\hbar\omega}{\gamma k}\right)^2}
% \right\},\nonumber
% \end{eqnarray}
% where $K_0$ and $K_1$ denote modified Bessel functions of the second kind, also referred to as MacDonald functions or modified Hankel functions \cite{abramowitz}.
% The result for the real part of $\Pi$ is given by
% \begin{eqnarray}\label{RePilin3}
% \mathrm{Re}\,\Pi_{aa}(k,\omega)
% &=&-\frac{1}{4\pi\beta\gamma^2}\,n_a\Lambda_a^2
% \left\{
% K_1\left(\frac{\beta\gamma k}{2}\right)\,
% \sum\limits_{j=0}^{\infty}I_{2j+1}\left(\frac{\beta\gamma k}{2}\right)
% U_{2j}\left(\frac{\hbar\omega}{\gamma k}\right)\right.\nonumber\\
% &&\hspace*{12ex}\left.
% +K_0\left(\frac{\beta\gamma k}{2}\right)\,
% \sum\limits_{j=1}^{\infty}2j\,I_{2j}\left(\frac{\beta\gamma k}{2}\right)
% T_{2j}\left(\frac{\hbar\omega}{\gamma k}\right)
% \right\}\,.\nonumber
% \end{eqnarray}

\section{Extension to moderate degeneracy}

So far, the analysis relied on the assumption of very weak degeneracy of the carriers which allows to assume Boltzmann distributions (\ref{fboltzmann}).
Now we look more closely at the distribution function. It reads for arbitrary degeneracy (Fermi distribution)
\begin{equation}\label{ffermi}
f_a(k)=\frac{1}{\exp\left(\frac{\hbar^2k^2}{2m_ak_{\rm B}T_a}-\mu_a\right)+1}\,,
\end{equation}
where $\mu_a$ is the chemical potential of species $a$.
Introducing the fugacity $z=$ e$^{\beta\mu_a}$ and abbreviating the Boltzmann factor
$b=\exp\left(-\frac{\hbar^2k^2}{2m_ak_{\rm B}T_a}\right)$ one can write
\begin{eqnarray}\label{ffermi1}
f_a(k)=\frac{1}{z^{-1}b^{-1}+1}=\frac{zb}{1+zb}=zb\left(1-zb+z^2b^2-z^3b^3+-...\right)\,,
\end{eqnarray}
where the last equality holds for $z<1$, i.e., for weak to moderate degeneracy. The Boltzmann factor to an arbitrary power $j$ reads
\begin{equation}
b^j=\exp\left(-j\frac{\hbar^2k^2}{2m_ak_{\rm B}T_a}\right)
=\exp\left(-\frac{\hbar^2k^2}{2m_ak_{\rm B}(T_a/j)}\right)\,,
\end{equation}
i.e., it corresponds to a Boltzmann factor with an effective temperature $T/j$. We get
\begin{eqnarray}\label{ffermi2}
f_a(k)&=&z\,\exp\left(-\frac{\hbar^2k^2}{2m_ak_{\rm B}T_a}\right)\sum\limits_{j=0}^{\infty}(-1)^jz^j\,\exp\left(-\frac{\hbar^2k^2}{2m_ak_{\rm B}(T_a/j)}\right)\nonumber\\
&=&\sum\limits_{j=1}^{\infty}(-1)^{j-1}z^j\,\exp\left(-\frac{\hbar^2k^2}{2m_ak_{\rm B}(T_a/j)}\right)\,.
\end{eqnarray}

The difference of distribution functions occurring in $\Pi$  (\ref{Pi}) can be calculated in every order of the expansion (\ref{ffermi2}) analogously to (\ref{fdiff}) (or in the Cartesian analogue leading to (\ref{Pi1-2d}) and (\ref{I1a-2d-cart-1}), respectively). Therefore, the calculation in each order is the same as presented in the previous sections. The result is a series for $\Pi$,
\begin{eqnarray}\label{Pi-reihe}
\Pi_{aa}^{\rm qc}(k,\omega;T_a)
=\frac{2}{n_a\Lambda_a^d}\sum\limits_{j=1}^{\infty}\frac{(-1)^{j-1}}{j^{d/2}}\,z^j\,\Pi_{aa}(k,\omega;T_a/j)\,,
\end{eqnarray}
where $\Pi_{aa}$ denotes the function in the weakly degenerate case derived in the previous sections. We should note here that this result is exact within the convergence radius of the series, i.e., for $z<1$, while its validity is restricted by the choice of approximation for the fugacity. The first few elements of the series with $j\ge 2$ may be regarded as \textit{quantum corrections} to the nondegenerate result $(j=1)$. We then can write down the quantum correction of the order $j$ for $\Pi$ as (i.e., shift the index by 1)
\begin{eqnarray}\label{Pi-qcj}
\Pi_{aa}^{\rm qc(j)}(k,\omega;T_a)=(-1)^j\frac{2}{n_a\Lambda_a^d}\frac{1}{(j+1)^{d/2}}\,z^{j+1}\,\Pi_{aa}(k,\omega;T_a/(j+1))\,.
%\Pi_{aa}^{\rm qc(2)}(k,\omega;T_a)=\frac{1}{3^{d/2}}\left(\frac{n_a\Lambda_a^d}{2}\right)^2\Pi_{aa}(k,\omega;T_a/3)\,.
\end{eqnarray}

%\section{Numerical illustration}

In order to illustrate the results, we consider in this section electrons and holes in bulk Cu$_2$O. For the fugacity we use the nondegenerate limit $z=n_a\Lambda_a^3/2$.
\begin{figure}[h]%
\begin{center}
\includegraphics*[width=0.49\textwidth]{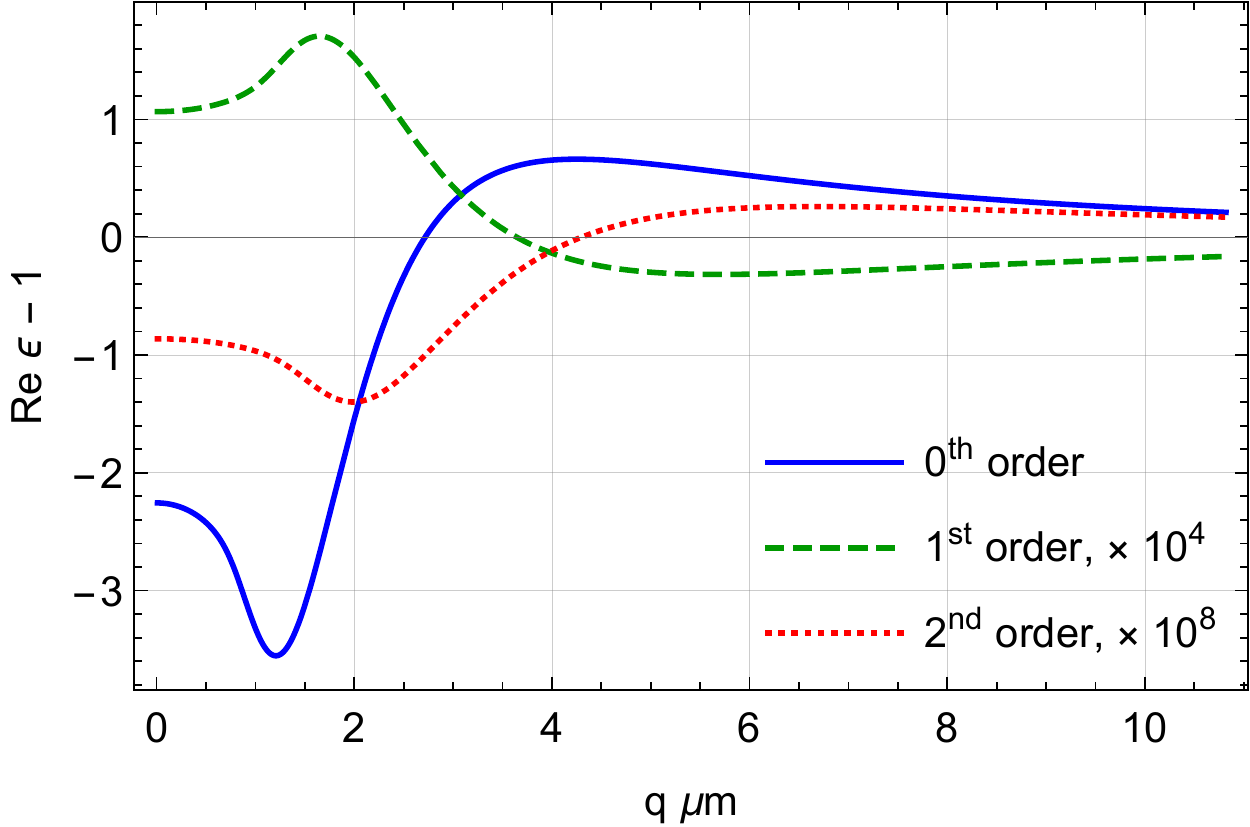}\hspace*{0.1cm}
\includegraphics*[width=0.49\textwidth]{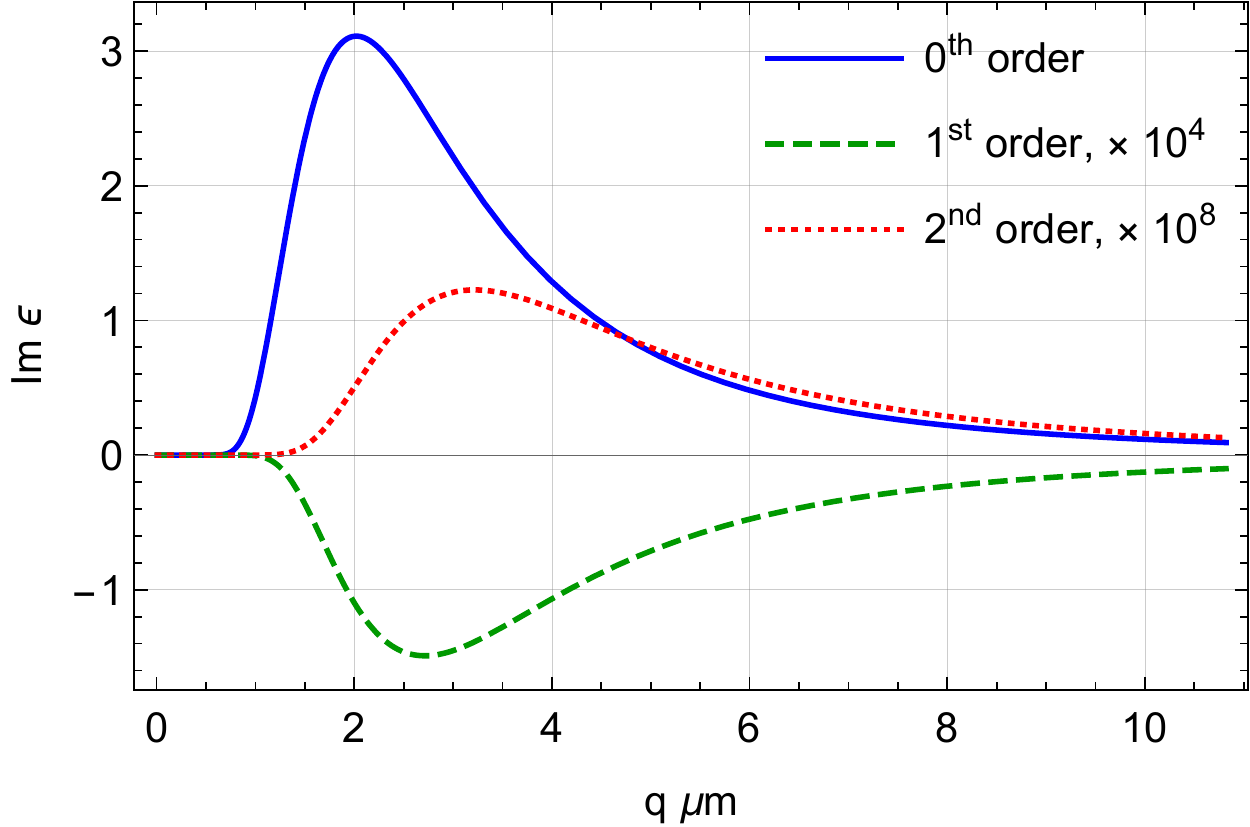}
\caption{Quantum correction terms of first and second order to the real (left panel) and imaginary (right panel) parts of the dielectric function of electrons and holes in bulk Cu$_2$O compared to the nondegenerate case \textit{vs.}\ wave number $q$ at frequency $\hbar\omega=15$ \textmu eV for temperature $T=2$ K and carrier density $n=10^{12}$ cm$^{-3}$. Note the magnification factors of the first and second order terms.
}
\label{fig:epsilon-qc-j}
\end{center}
\end{figure}
Figure \ref{fig:epsilon-qc-j} shows the first two quantum correction terms as a function of the wave number $q$. They are, obviously, tiny for the chosen parameters even though those can be regarded as upper (density) and lower (temperature) bounds, respectively, being relevant for (current) experiments investigating  Rydberg excitons \cite{nature2014,heckoetter2018}.

However, there are situations where the quantum degeneracy is much higher, e.g., in the experiments attempting to prove the existence of an excitonic Bose-Einstein condensate in bulk Cu$_2$O. There, electron-hole densities around $10^{16}$ cm$^{-3}$ have been generated by the optical excitation \cite{stolz2012}.

\begin{figure}[h]%
\begin{center}
\includegraphics*[width=0.49\textwidth]{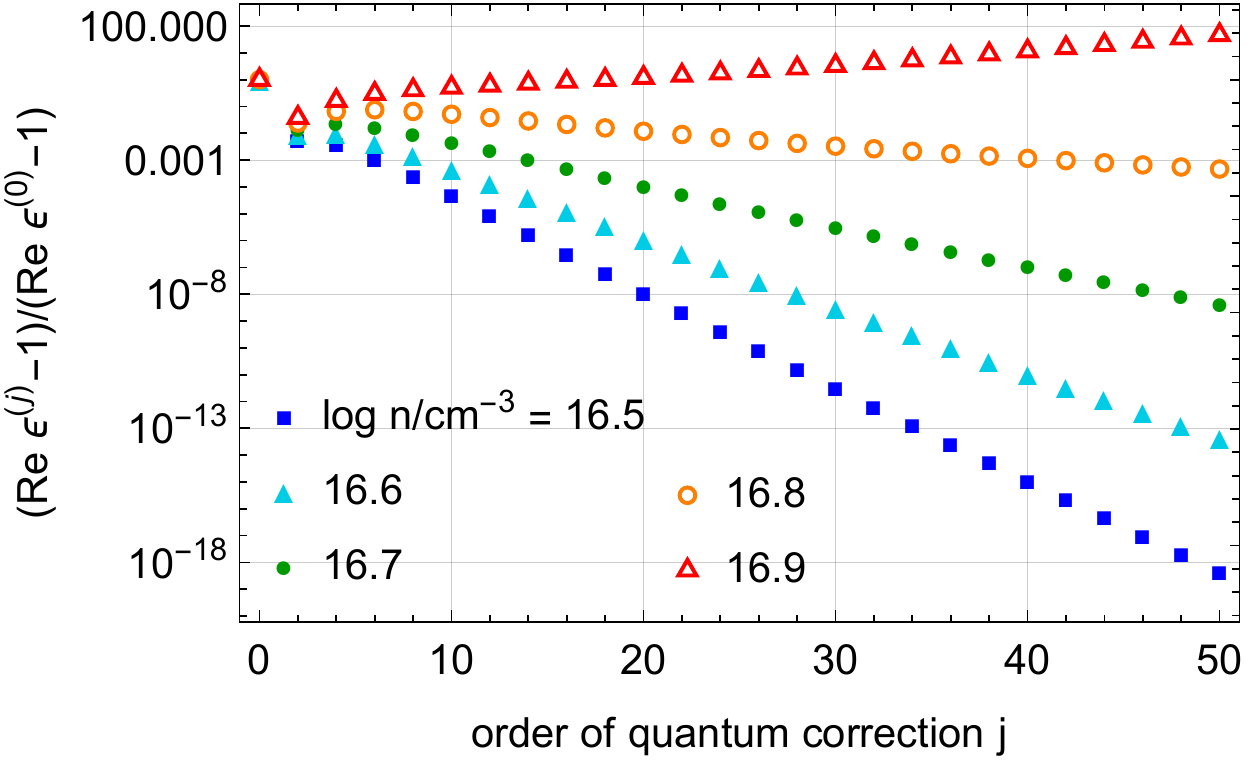}\hspace*{0.1cm}
\includegraphics*[width=0.49\textwidth]{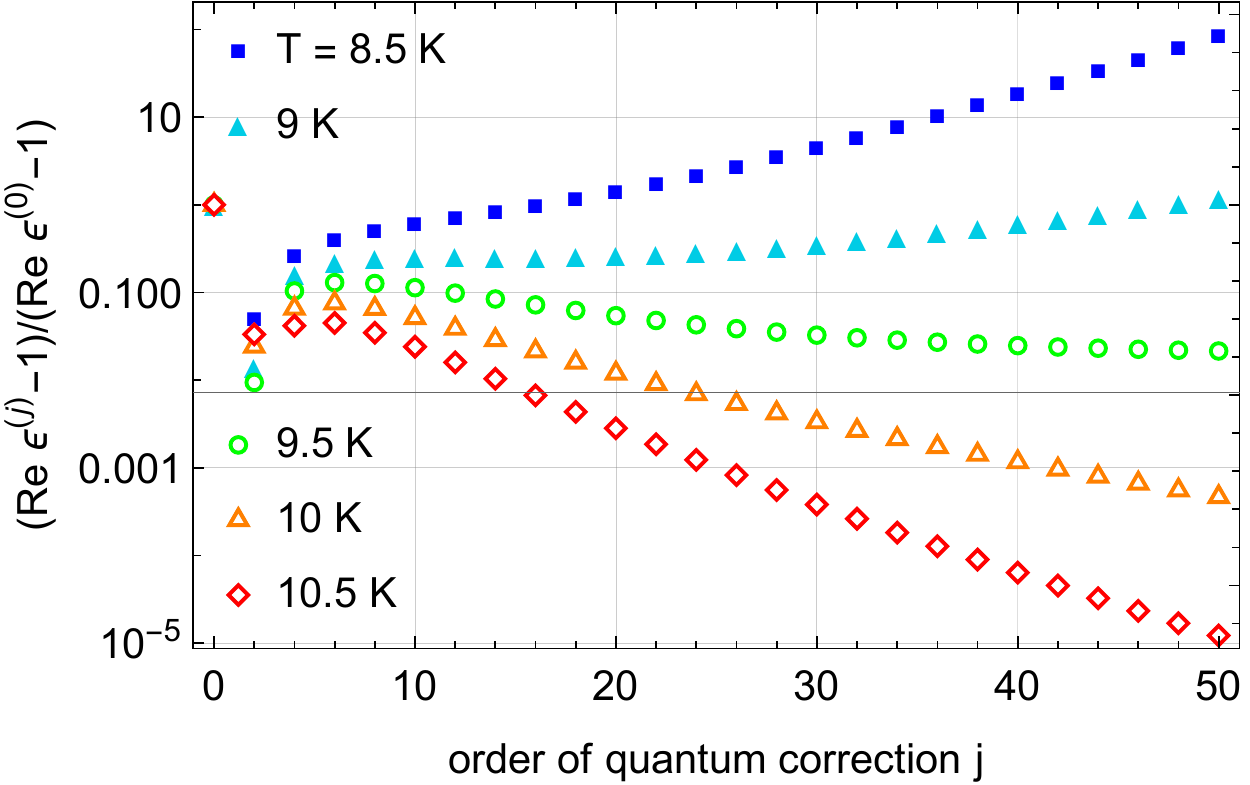}
\caption{Quantum correction term of $j^{\rm th}$ order to the real part of the dielectric function normalized to the nondegenerate term \textit{vs.} order $j$ for a temperature of $T=10$ K and several carrier densities (left panel) and for log $n/$cm$^{-3}=16.5$ and several temperatures (right panel).
}
\label{fig:epsilon-qc-j-rho}
\end{center}
\end{figure}

For the further analysis, we restrict ourselves to the real part and look at the magnitude of the quantum correction terms at fixed wave number and frequency relative to the nondegenerate case, if not given explicitly, at a wave number of $q=2$ \textmu m$^{-1}$ and a frequency of $\hbar\omega=15$ \textmu eV. Figure \ref{fig:epsilon-qc-j-rho} shows $(\mathrm{Re}\,\varepsilon^{(j)}-1)/(\mathrm{Re}\,\varepsilon^{(0)}-1)$ depending on the order $j$ for several densities and temperatures. For a convergent series, the terms have at least to decrease with increasing order which is the case (at $T=10$ K) only for log $n/$cm$^{-3}\le 16.8$ (left panel) and (at log $n/$cm$^{-3}=16.8$) only for $T\ge 10$~K (right panel). Indeed, the border of $n\Lambda^3=1/2$ lies for the (lighter) holes with $T=10$ K just at  log~$n/$cm$^{-3}= 16.8234$ and with log $n/$cm$^{-3}=16.8$ just at $T=9.6472$ K. For densities or temperatures beyond that border, the series (\ref{Pi-reihe}) is not convergent and the terms (\ref{Pi-qcj}) have no physical interpretation.

\begin{figure}[h]%
\begin{center}
\includegraphics*[width=0.49\textwidth]{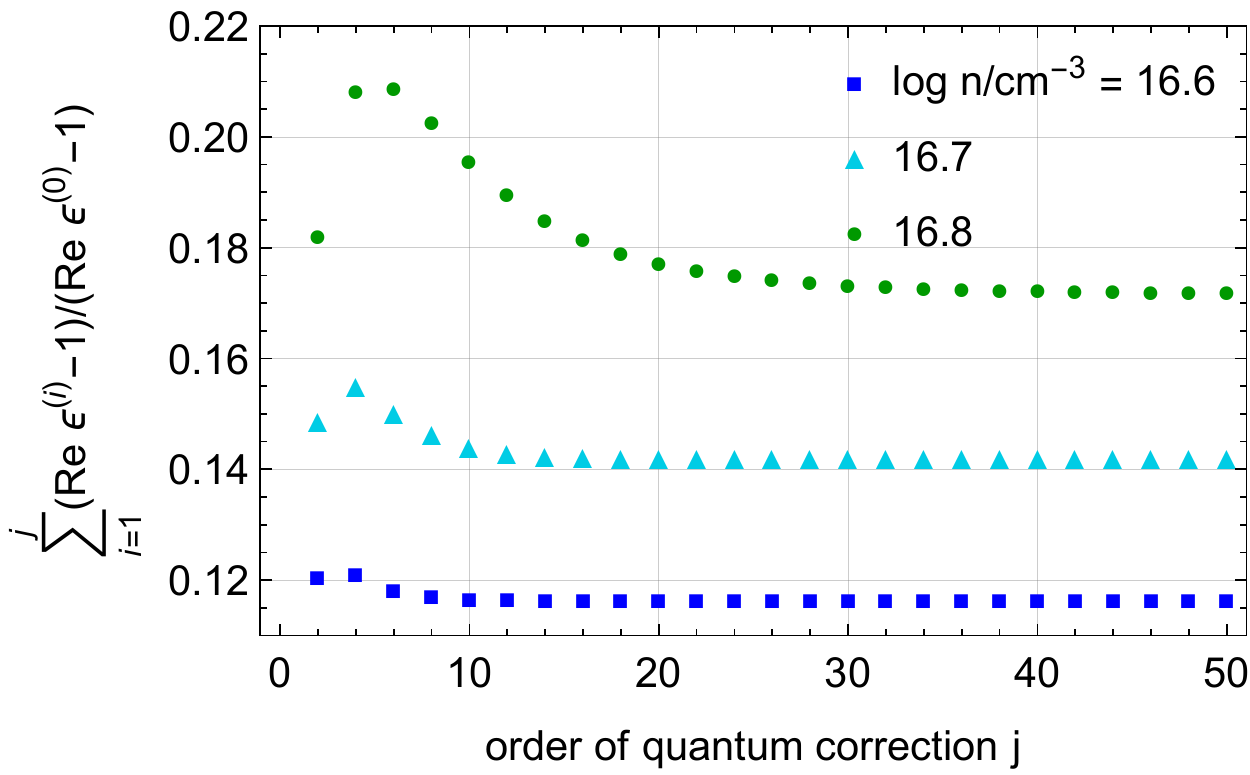}\hspace*{0.1cm}
\includegraphics*[width=0.49\textwidth]{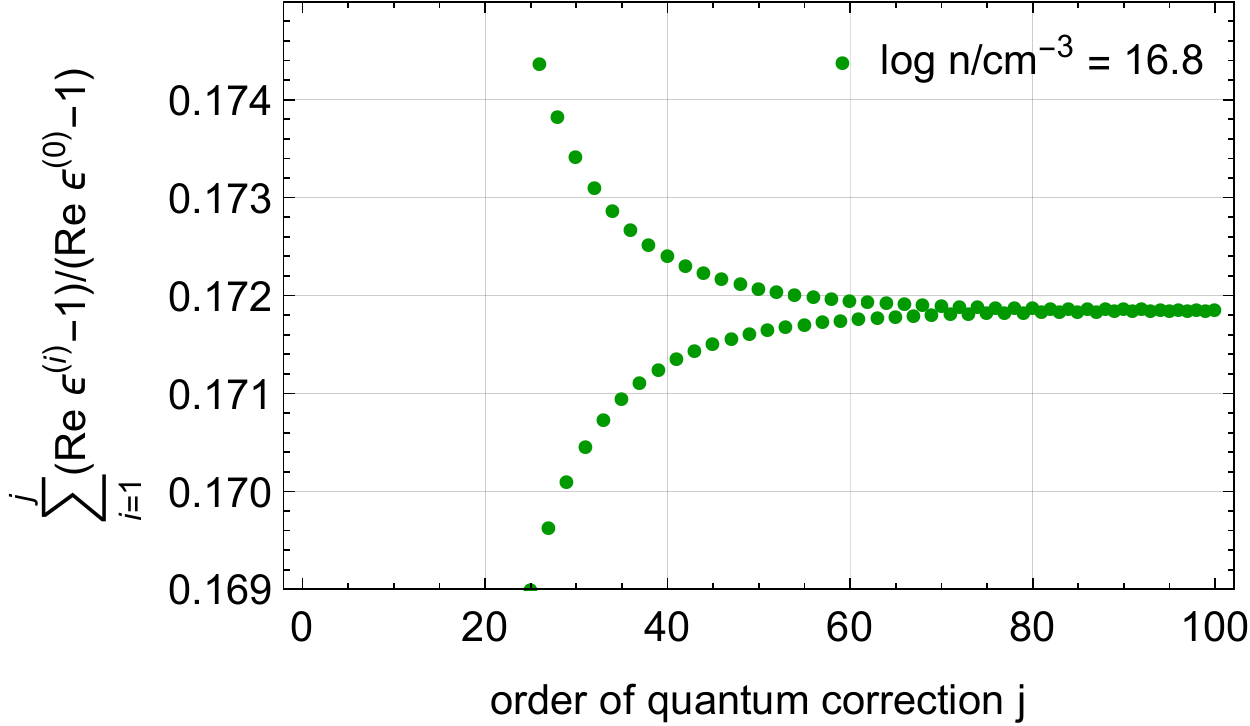}
\caption{Sum of quantum correction terms up to $j^{\rm th}$ order to the real part of the dielectric function normalized to the nondegenerate term \textit{vs.} order $j$ for a temperature of $T=10$ K and several carrier densities (left panel). Same quantity only for the highest density, but up to $j=100$ (right panel).
}
\label{fig:epsilon-qc-sum50-rho}
\end{center}
\end{figure}

The left panel of figure \ref{fig:epsilon-qc-sum50-rho} shows the convergence of the series. It is slower at the border of the covergence area (see also right panel), however, even summing up 100 terms is numerically still quite feasible.

\begin{figure}[htb]%
\begin{center}
\includegraphics*[width=0.49\textwidth]{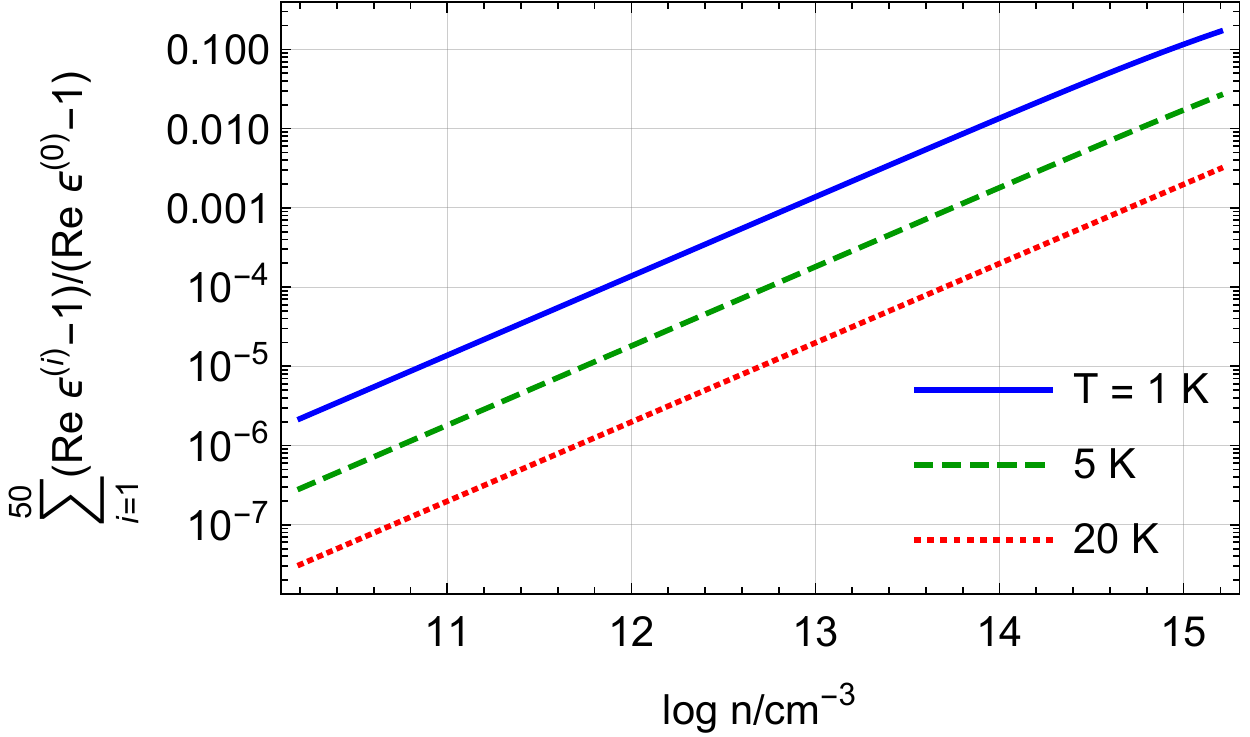}\hspace*{0.1cm}
\includegraphics*[width=0.49\textwidth]{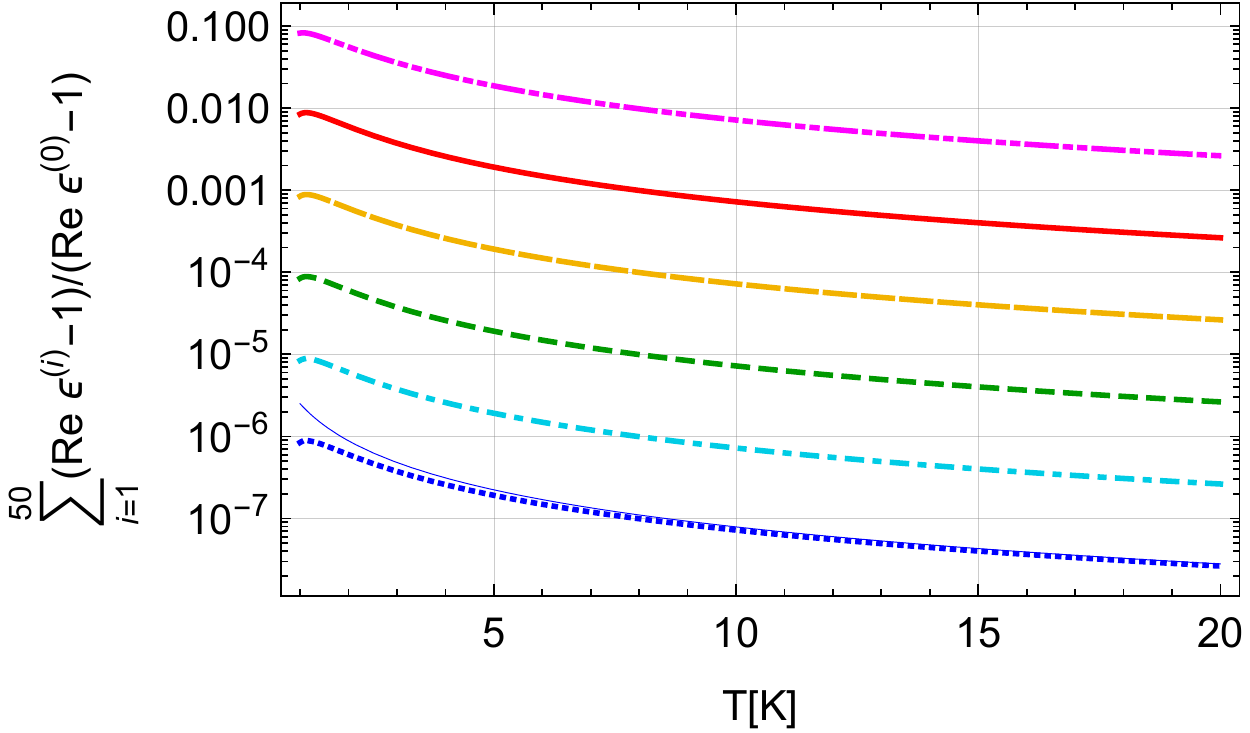}
\caption{Sum of quantum correction terms up to $50^{\rm th}$ order to the real part of the dielectric function normalized to the nondegenerate term \textit{vs.} carrier density at a frequency of $\hbar\omega=5$~\textmu eV for several temperatures (left panel) and \textit{vs.}\ temperature for several densities (right panel; from bottom to top: log $n/$cm$^{-3}=10, 11, 12, 13, 14, 15$). The thin solid blue line in the right-hand panel gives a $T^{-3/2}$ law.
}
\label{fig:epsilon-qc-sum50-rho-T}
\end{center}
\end{figure}

Finally, we consider the sum of quantum correction terms up to $50^{\rm th}$ order to the real part of the dielectric function normalized to the nondegenerate term as a function of the particle density for several temperatures and vice versa (figure \ref{fig:epsilon-qc-sum50-rho-T}). While the density dependence is obviously $\propto n$, the temperature dependence is $\propto T^{-3/2}$, see dashed line in the right panel.

\begin{figure}[htb]%
\begin{center}
\includegraphics*[width=0.49\textwidth]{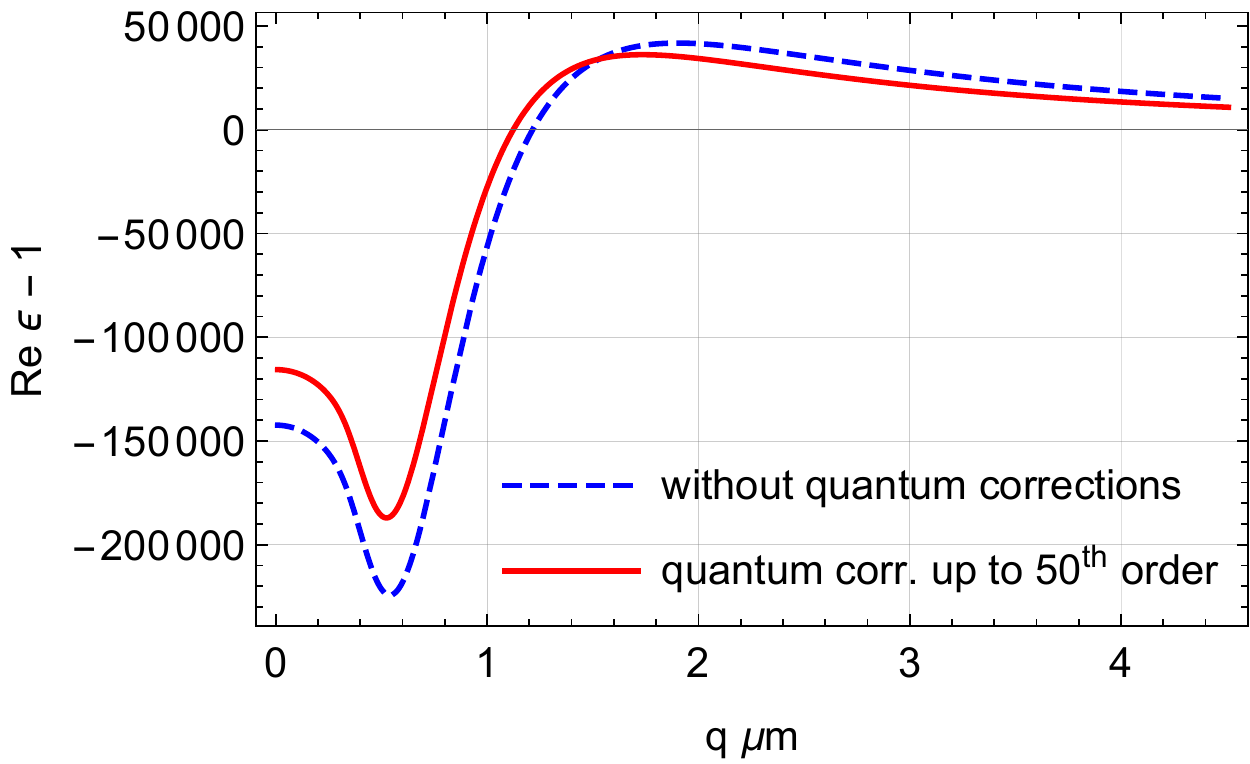}\hspace*{0.1cm}
\includegraphics*[width=0.49\textwidth]{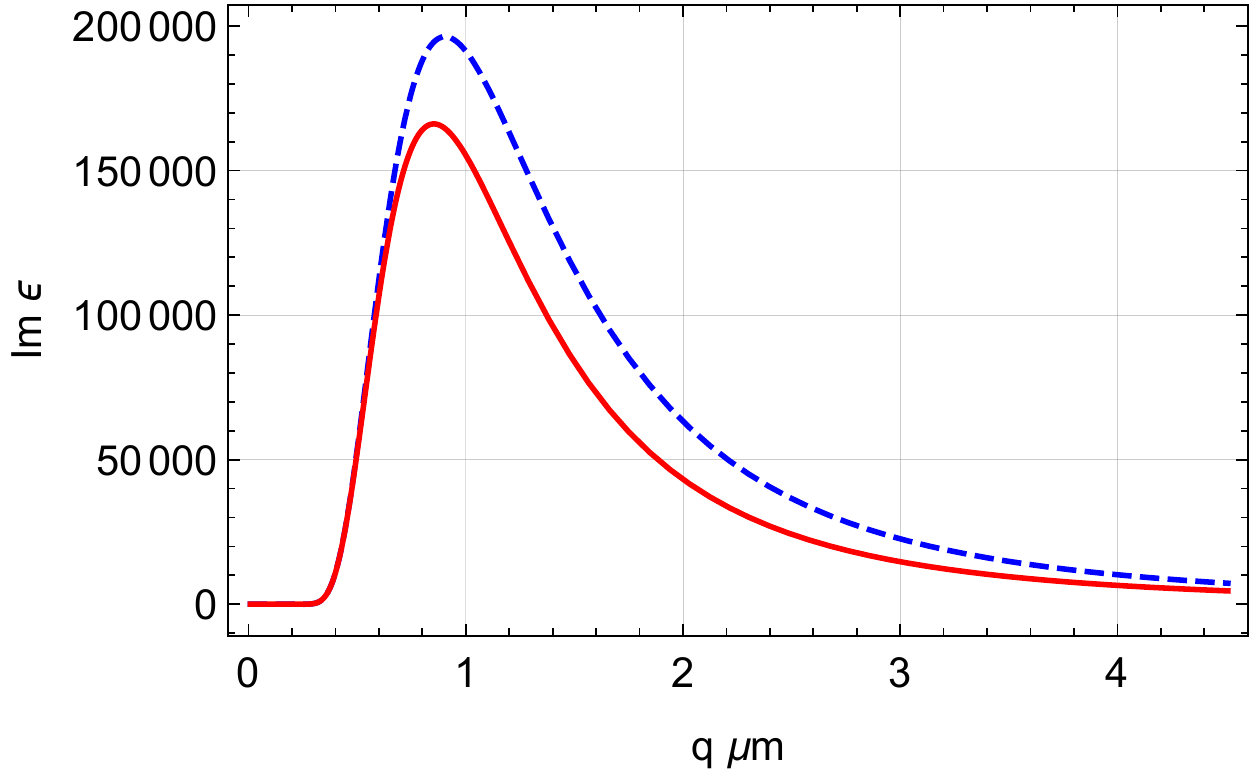}
\caption{Real part (left panel) and imaginary part (right panel) of the dielectric function \textit{vs.}\ wave number for a carrier density of log $n/$cm$^{-3}=16.8$ and a temperature of 10 K. Comparison of nondegenerate term (dashed blue) and quantum corrected function (sum of nondegenerate term and quantum corrections up to 50$^{\rm th}$ order; red).
}
\label{fig:epsilon-comp}
\end{center}
\end{figure}

Figure \ref{fig:epsilon-comp} illustrates the effect of quantum corrections on the dielectric function by comparing the weakly degenerate limit and the function including quantum corrections for a system with quantum degeneracy (of the holes) of $\frac{1}{2}n\Lambda_{\rm h}^3=0.95$.

\section{Conclusions and outlook}

We have derived analytical (RPA) results for the dielectric response of an electron-hole plasma in the weakly-degenerate case and demonstrated that the well-tried result for bulk systems \cite{klimontovich74} keeps its form also for lower-dimensional structures. Moreover, it is even in the more complicated case of linear carrier dispersion, realized, e.g., in graphene and many topological insulators, possible to derive a result for the polarization function for excited states in an analytical form (see \ref{app:lindisp}). Since that function determines, in particular, also the plasmonic properties of bulk and lower-dimensional semiconductors, its knowledge in analytical form can be expected of great usefulness for the calculation of these properties.

In order to generalize the results beyond the weakly-degenerate case, we have established a method which expands the polarization function in a series with respect to the fugacity $z=\mathrm{e}^{\beta\mu}$. While the whole series covers the region of weak and moderate degeneracies up to $n\Lambda^d/2=1$, its first terms can be regarded as \textit{quantum corrections} to the result in the nondegenerate limit.

For the parameters relevant in the Rydberg exciton experiments in bulk cuprous oxide \cite{nature2014,heckoetter2018} (ultralow carrier densities of $n\lesssim 10^{12}$ cm$^{-3}$), the nondegenerate limit for $\varepsilon$ is a very good approximation, and the quantum correction terms are negligible. Obviously, this is not the case for higher densities as in experiments searching for an excitonic Bose-Einstein condensate \cite{stolz2012}. Moreover, one can expect that quantum corrections will play a much more important role in lower-dimensional systems, particularly also in those with Dirac cone functionality.

\ack
%\section{Acknowledgements}

%We would like to thank ...
D.\ S.\ gratefully acknowledges support by the Deutsche Forschungsgemeinschaft (project number SE 2885/1-1).

\appendix

\section{Derivation of the effective quasi-two-dimensional Coulomb interaction}\label{veff2d}

Starting point is the effective Coulomb potential between two carriers (electrons and holes) in a quantum well \cite{girndt1997},
\begin{eqnarray}\label{veh1}
V_{ab}(q)=\frac{e_ae_b}{2\epsilon_0\epsilon_{\rm b,w}q}\int\limits_{-\infty}^{\infty}\mathrm{d}z\int\limits_{-\infty}^{\infty}\mathrm{d}z'
\left|\phi_a(z)\right|^2\left|\phi_b(z')\right|^2\,\exp\left(-q|z-z'|\right).
\end{eqnarray}
Here, $\phi_{a/b}$ are the wave functions of the motion in confinement- ($z$-)direction, $\epsilon_{\rm b,w}$ is the background dielectric constant in the well, and $e_{\rm e/h}=\mp e$.

The wave functions are those of a particle confined in a one-dimensional quantum well. They read for the case of even (odd) states
 \cite{bastard1988}
\begin{eqnarray}\label{wf1}
\phi_a(z)=\left\{
\begin{array}{ll}
\eta_a B_a\,\exp\left[\kappa_a\left(z+\frac{d}{2}\right)\right]\,, & -\infty<z\le -\frac{d}{2}\\
A_a\,cs(k_az)\,, &  -\frac{d}{2}\le z\le \frac{d}{2}\\
B_a\,\exp\left[-\kappa_a\left(z-\frac{d}{2}\right)\right]\,, & \frac{d}{2}\le z<\infty
\end{array}
\right.
\end{eqnarray}
with $a= $ e,h, $k_a=\left(\frac{2m_{a,\mathrm{w}}}{\hbar^2}E\right)^{1/2}$ and $\kappa_a=\left(\frac{2m_{a,\mathrm{b}}}{\hbar^2}(V_{0,a}-E)\right)^{1/2}$, $m_{a,\mathrm{w}}$ and $m_{a,\mathrm{b}}$ being the masses of carrier species $a$ in the well and in the barriers, respectively, $V_{0,a}$ the barrier height, and $E$ the energy of the confined particle.
In order to account for even and odd states of electrons and holes, we introduced the factors $\eta_a=\pm 1$ for even (odd) states ($a=$ e,h) and the functions
\begin{eqnarray}
ct(\theta_a)&=&\left\{
\begin{array}{l}
\cot\,\theta_a\\
\tan\,\theta_a\,,
\end{array}
\right.
\quad
tc(\theta_a)=\left\{
\begin{array}{l}
\tan\,\theta_a\\
\cot\,\theta_a\,,
\end{array}
\right.
\nonumber\\
cs(\theta_a)&=&\left\{
\begin{array}{l}
\cos\,\theta_a\\
\sin\,\theta_a\,,
\end{array}
\right.
\quad
sc(\theta_a)=\left\{
\begin{array}{l}
\sin\,\theta_a\\
\cos\,\theta_a
\end{array}
\right.
\end{eqnarray}
(upper (lower) functions for even (odd) states).

In order to determine the normalization constants $A_a$ and $B_a$, we apply the normalization condition of the wave functions,
\begin{eqnarray}\label{norm1}
1=\int\limits_{-\infty}^{\infty}\mathrm{d}z\,\left|\phi_a(z)\right|^2&=&
B_a^2\int\limits_{-\infty}^{-d/2}\mathrm{d}z\,\exp\left[2\kappa_a\left(z+\frac{d}{2}\right)\right]
+A_a^2\int\limits_{-d/2}^{d/2}\mathrm{d}z\,cs^2(k_az)\nonumber\\
&&+B_a^2\int\limits_{d/2}^{-\infty}\mathrm{d}z\,\exp\left[-2\kappa_a\left(z-\frac{d}{2}\right)\right]\nonumber\\
&=&\frac{B_a^2}{\kappa_a}+\frac{A_a^2d}{2}\left(1+\eta_a\frac{\sin k_ad}{k_ad}\right)\,.
\end{eqnarray}

Further information can be obtained by making use of the boundary conditions at $z=\pm d/2$, i.e., of the continuity of wave functions and particle fluxes \cite{fox2006}. One gets
\begin{eqnarray}\label{boundarycond}
B_a=A_a\,cs\left(\frac{k_ad}{2}\right)\quad\mbox{and}\quad \frac{B_a\kappa_a}{m_{a,\mathrm{b}}}=\eta_a\frac{A_ak_a}{m_{a,\mathrm{w}}}sc\left(\frac{k_ad}{2}\right)\,.
\end{eqnarray}
The first relation already allows to eliminate one of the two constants. Even more importantly, both relations together lead to
\begin{eqnarray}\label{energycond}
\eta_a\frac{m_{a,\mathrm{w}}\kappa_a}{m_{a,\mathrm{b}}k_a}=tc\left(\frac{k_ad}{2}\right)\,,
\end{eqnarray}
i.e., a condition for the possible energies of the particle's motion in $z$-direction. The solutions can be illustrated most easily by abbreviating
\begin{eqnarray}\label{theta}
\theta_a=\frac{k_ad}{2}\,,\quad\theta_{a,0}=\frac{k_{a,0}d}{2}\,,\quad k_{a,0}=\left(\frac{2m_{a,\mathrm{w}}}{\hbar^2}V_{0,a}\right)^{1/2}\,,
\quad\alpha_a=\frac{m_{a,\mathrm{w}}}{m_{a,\mathrm{b}}}
\end{eqnarray}
which renders (\ref{energycond}) into
\begin{eqnarray}\label{energycond1}
\eta_a\sqrt{\alpha_a}\left(\frac{\theta_{a,0}^2}{\theta_a^2}-1\right)^{1/2}=tc(\theta_a)\,.
\end{eqnarray}

We proceed now with the further evaluation of (\ref{norm1}). Inserting (\ref{boundarycond})--(\ref{theta}) one obtains
\begin{eqnarray}\label{norm2}
1=
%\frac{B_a^2}{\kappa_a}+\frac{A_a^2d}{2}\left(1+\eta_a\frac{\sin k_ad}{k_ad}\right)\nonumber\\
\frac{A_a^2d}{2}\left[1+\eta_a\frac{\alpha_a}{\theta_a}ct(\theta_a)+\eta_a\frac{1-\alpha_a}{\theta_a}\cos\,\theta_a\sin\,\theta_a
\right]
\,,
\end{eqnarray}
i.e., the normalization constant $A_a$ follows to be
\begin{eqnarray}\label{norm3}
A_a=\left(\frac{2}{d}\right)^{1/2}\frac{1}{\left(1+\eta_a\frac{\alpha_a}{\theta_a}ct(\theta_a)+\eta_a\frac{1-\alpha_a}{\theta_a}\cos\,\theta_a\sin\,\theta_a\right)^{1/2}}\,.
\end{eqnarray}

Now we go back to (\ref{veh1}). Inserting the wave function (\ref{wf1}), the potential consists of three parts,
\begin{eqnarray}\label{veh2}
V_{ab}(q)=\frac{e_ae_b}{2\epsilon_0\epsilon_{\rm b,w}q}(I_1+I_2+I_3)
\end{eqnarray}
with
\begin{eqnarray}
\fl I_1&=&B_a^2B_b^2\int\limits_{-\infty}^{-d/2}\mathrm{d}z\int\limits_{-\infty}^{-d/2}\mathrm{d}z'\,
\exp\left[2\kappa_a\left(z+\frac{d}{2}\right)\right]\,\exp\left[2\kappa_b\left(z'+\frac{d}{2}\right)\right]\,\exp\left(-q|z-z'|\right)\nonumber\\
\fl&=&B_a^2B_b^2\left\{\int\limits_{-\infty}^{-d/2}\mathrm{d}z'\int\limits_{-\infty}^{z'}\mathrm{d}z\,
\exp\left[2\kappa_a\left(z+\frac{d}{2}\right)\right]\,\exp\left(2\kappa_b\left(z'+\frac{d}{2}\right)\right]\,\exp\left[-q(z'-z)\right]\right.
\nonumber\\
\fl&&\hspace*{7ex}
+\int\limits_{-\infty}^{-d/2}\mathrm{d}z'\int\limits_{z'}^{-d/2}\mathrm{d}z\,
\exp\left[2\kappa_a\left(z+\frac{d}{2}\right)\right]\,\exp\left[2\kappa_b\left(z'+\frac{d}{2}\right)\right]\nonumber\\
\fl&&\hspace*{7ex}\exp\left[-q(z-z')\right]
\Bigg\}\nonumber\\
\fl&=&B_a^2B_b^2\,\frac{\kappa_a+\kappa_b+q}{(\kappa_a+\kappa_b)(2\kappa_a+q)(2\kappa_b+q)}\,.
\end{eqnarray}
Using the relation (\ref{energycond}) and the abbreviations (\ref{theta}) (furthermore $Q=qd/2$), and inserting (\ref{boundarycond}) and (\ref{norm3}), one arrives at
\begin{eqnarray}
\fl I_1=\frac{\alpha_a\alpha_b\,cs^2(\theta_a)\,cs^2(\theta_b)}
{\left(1+\eta_a\frac{\alpha_a}{\theta_a}\,ct(\theta_a)+\eta_a\frac{1-\alpha_a}{\theta_a}\cos\,\theta_a\sin\,\theta_a\right)
\left(1+\eta_b\frac{\alpha_b}{\theta_b}\,ct(\theta_b)+\eta_b\frac{1-\alpha_b}{\theta_b}\cos\,\theta_b\sin\,\theta_b\right)}\nonumber\\
\fl\hspace*{5ex}\times
\frac{\eta_a\,\alpha_b\,\theta_a\,tc(\theta_a)+\eta_b\,\alpha_a\,\theta_b\,tc(\theta_b)+\alpha_a\alpha_b Q}
{\left(\eta_a\,\alpha_b\,\theta_a\,tc(\theta_a)+\eta_b\,\alpha_a\,\theta_b\,tc(\theta_b)\right)
\left(2\eta_a\,\theta_a\,tc(\theta_a)+\alpha_a Q\right)\left(2\eta_b\,\theta_b\,tc(\theta_b)+\alpha_b Q\right)}\,.
\end{eqnarray}
The calculation of $I_3$ runs analogously with the result $I_3=I_1$.

For $I_2$ one gets
\begin{eqnarray}
\fl I_2&=&A_a^2A_b^2\int\limits_{-d/2}^{d/2}\mathrm{d}z\int\limits_{-d/2}^{d/2}\mathrm{d}z'\,
cs^2(k_az)\,cs^2(k_bz')\,\exp\left(-q|z-z'|\right)\nonumber\\
\fl &=&\frac{d}{2q}A_a^2A_b^2\left\{
1+\eta_a\frac{\sin(k_a d)}{k_a d}+\eta_b\frac{q^2}{4k_b^2+q^2}\left[
\frac{\sin(k_b d)}{k_b d}+\eta_a\frac{\sin((k_a+k_b)d)}{2(k_a+k_b)d}\right.\right.\nonumber\\
\fl &&\left.+\eta_a\frac{\sin((k_a-k_b)d)}{2(k_a-k_b)d}
\right]\nonumber\\
\fl &&-\frac{1}{qd}\left[1-\mathrm{e}^{-qd}+\eta_a\frac{q^2}{4k_a^2+q^2}
\left(\cos(k_a d)\left(1-\mathrm{e}^{-qd}\right)
+\frac{2k_a}{q}\sin(k_a d)\left(1+\mathrm{e}^{-qd}\right)\right)
\right]\nonumber\\
\fl &&\hspace*{3ex}\left.\times
\left[1+\eta_b\frac{q^2}{4k_b^2+q^2}
\left(\cos(k_b d)-\frac{2k_b}{q}\sin(k_b d)\right)
\right]
\right\}
\,.
\end{eqnarray}
Using now again the relations (\ref{energycond}) and (\ref{norm3}) and the abbreviations (\ref{theta}) and $Q=qd/2$, one finally arrives at
\begin{eqnarray}
\fl I_2=\frac{1}{Q}
\frac{1}
{\left(1+\eta_a\frac{\alpha_a}{\theta_a}ct(\theta_a)+\eta_a\frac{1-\alpha_a}{\theta_a}\cos\,\theta_a\sin\,\theta_a\right)
\left(1+\eta_b\frac{\alpha_b}{\theta_b}ct(\theta_b)+\eta_b\frac{1-\alpha_b}{\theta_b}\cos\,\theta_b\sin\,\theta_b\right)}\nonumber\\
\fl\hspace*{4ex} \times
\left\{
1+\eta_a\frac{\sin(2\theta_a)}{2\theta_a}+\eta_b\frac{Q^2}{4\theta_b^2+Q^2}\left[
\frac{\sin(2\theta_b)}{2\theta_b}+\eta_a\frac{\sin(2(\theta_a+\theta_b))}{4(\theta_a+\theta_b)}
+\eta_a\frac{\sin(2(\theta_a-\theta_b))}{4(\theta_a-\theta_b)}
\right]\right.\nonumber\\
\fl\hspace*{4ex} -\frac{1}{2Q}\left[1-\mathrm{e}^{-2Q}+\eta_a\frac{Q^2}{4\theta_a^2+Q^2}
\left(\cos(2\theta_a)\left(1-\mathrm{e}^{-2Q}\right)
+\frac{2\theta_a}{Q}\sin(2\theta_a)\left(1+\mathrm{e}^{-2Q}\right)\right)
\right]\nonumber\\
\fl \hspace*{4ex}\left.\times
\left[1+\eta_b\frac{Q^2}{4\theta_b^2+Q^2}
\left(\cos(2\theta_b)-\frac{2\theta_b}{Q}\sin(2\theta_b)\right)
\right]
\right\}
\,.
\end{eqnarray}

We insert the results for $I_1=I_3$ and $I_2$ into (\ref{veh2}) and obtain for the effective Coulomb interaction in a quantum well
\begin{eqnarray}\label{veh3}
\fl V_{ab}(Q)=\frac{de_ae_b}{4\epsilon_0\epsilon_{\rm b,w}Q}\nonumber\\
\fl\hspace*{2ex} \times\frac{1}
{\left(1+\eta_a\frac{\alpha_a}{\theta_a}ct(\theta_a)+\eta_a\frac{1-\alpha_a}{\theta_a}\cos\,\theta_a\sin\,\theta_a\right)
\left(1+\eta_b\frac{\alpha_b}{\theta_b}ct(\theta_b)+\eta_b\frac{1-\alpha_b}{\theta_b}\cos\,\theta_b\sin\,\theta_b\right)}\nonumber\\
\fl\hspace*{2ex} \times
\left(
\frac{\alpha_a\alpha_b\,cs^2(\theta_a)\,cs^2(\theta_b)\,(\eta_a\,\alpha_b\,\theta_a\,tc(\theta_a)+\eta_b\,\alpha_a\,\theta_b\,tc(\theta_b)+\alpha_a\alpha_b Q)}
{\left(\eta_a\,\alpha_b\,\theta_a\,tc(\theta_a)+\eta_b\,\alpha_a\,\theta_b\,tc(\theta_b)\right)
\left(2\eta_a\,\theta_a\,tc(\theta_a)+\alpha_a Q\right)\left(2\eta_b\,\theta_b\,tc(\theta_b)+\alpha_b Q\right)}
\right.\nonumber\\
\fl\hspace*{2ex} +\frac{1}{Q}
\left\{
1+\eta_a\frac{\sin(2\theta_a)}{2\theta_a}+\eta_b\frac{Q^2}{4\theta_b^2+Q^2}\left[
\frac{\sin(2\theta_b)}{2\theta_b}+\eta_a\frac{\sin(2(\theta_a+\theta_b))}{4(\theta_a+\theta_b)}
+\eta_a\frac{\sin(2(\theta_a-\theta_b))}{4(\theta_a-\theta_b)}
\right]\right.\nonumber\\
\fl\hspace*{2ex} -\frac{1}{2Q}\left[1-\mathrm{e}^{-2Q}+\eta_a\frac{Q^2}{4\theta_a^2+Q^2}
\left(\cos(2\theta_a)\left(1-\mathrm{e}^{-2Q}\right)
+\frac{2\theta_a}{Q}\sin(2\theta_a)\left(1+\mathrm{e}^{-2Q}\right)\right)
\right]\nonumber\\
\fl \hspace*{2ex}\left.\left.\times
\left[1+\eta_b\frac{Q^2}{4\theta_b^2+Q^2}
\left(\cos(2\theta_b)-\frac{2\theta_b}{Q}\sin(2\theta_b)\right)
\right]
\right\}
\right)
\,.
\end{eqnarray}

For illustration we consider the case of GaAs quantum wells embedded in Al$_x$Ga$_{1-x}$As barriers. This system has the following parameters \cite{ioffe,belov2019}: Electron masses $m_{\rm e,w}=0.063\,m_0$ and $m_{\rm e,b}=(0.063+0.083x)\,m_0$, hole masses $m_{\rm h,w}=0.51\,m_0$ and $m_{\rm h,b}=(0.51+0.25x)\,m_0$ (i.e., $\alpha_{\rm e}=0.717$ and $\alpha_{\rm h}=0.872$), dielectric constants $\epsilon_{\rm b,w}=12.90$ and $\epsilon_{\rm b,b}=12.90-2.84x$, and energy gap mismatch $\Delta E_{\rm g}=365.5$ meV (which splits onto electrons and holes like 0.65/0.35 so that $V_{0,\mathrm{e}}=237.575$ meV and  $V_{0,\mathrm{h}}=127.925$ meV). We use in the following $x=0.3$ and a well width of $d=20$ nm.

\begin{figure}[h]
\begin{center}
\includegraphics*[width=0.55\textwidth]{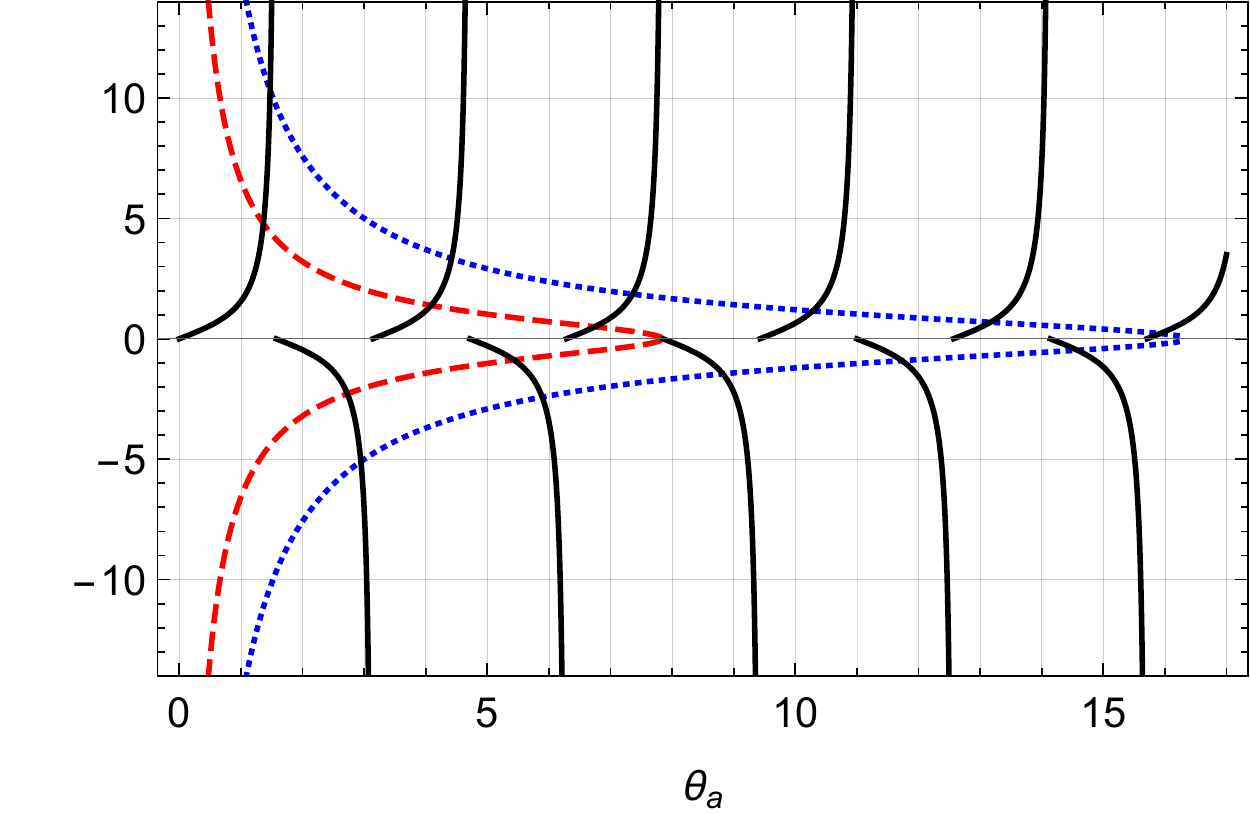}
\caption{Graphical solution of (\ref{energycond1}) for even (above the abscissa) and odd states (below the abscissa): r.h.s. (black solid lines) and l.h.s. for electrons (red dashed), and holes (blue dotted).
}
\label{fig:theta-gaas}
\end{center}
\end{figure}
Figure \ref{fig:theta-gaas} shows the graphical solution of (\ref{energycond1}). The electrons exhibit three even and two odd bound states, the holes exhibit six even and five odd bound states. Remember that $d=20$ nm here. The number of bound states increases with increasing well width.

\begin{figure}[h]
\begin{center}
\includegraphics*[width=0.495\textwidth]{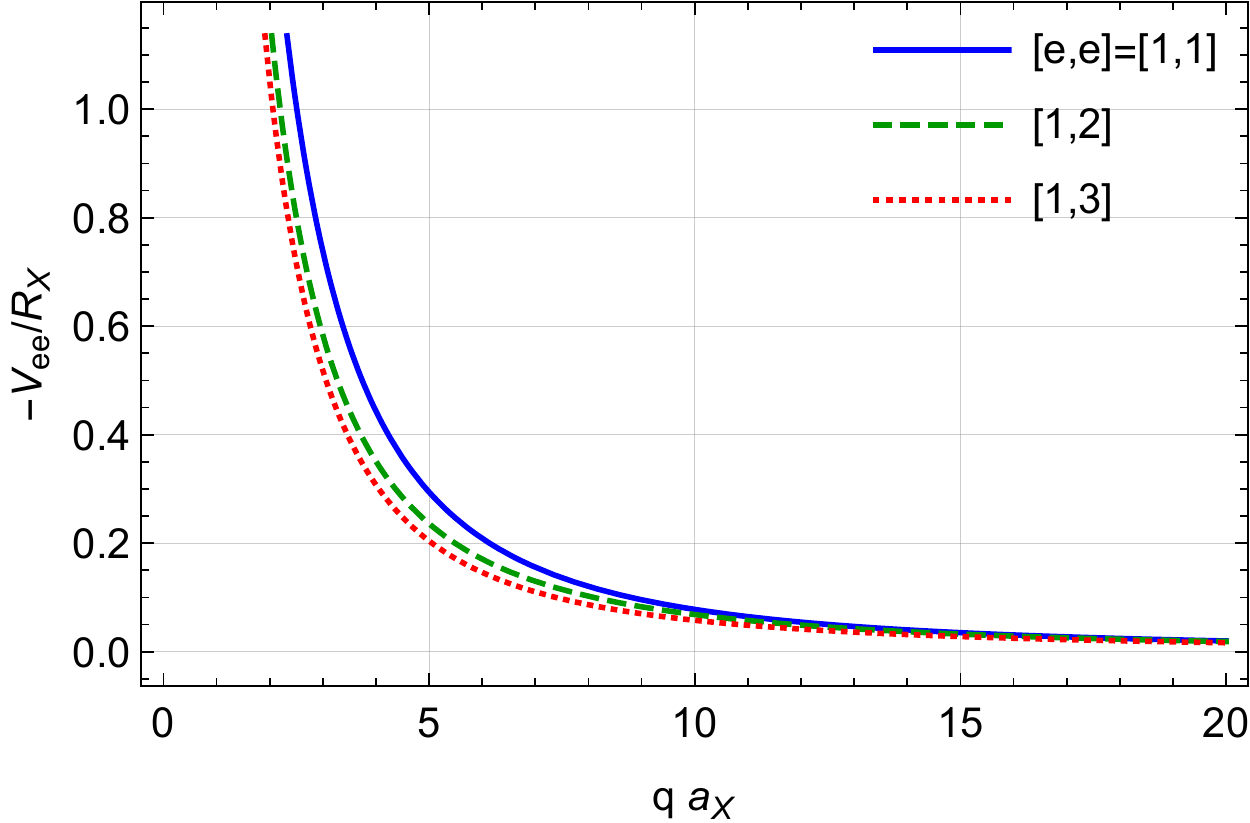}\hspace*{0.1cm}
\includegraphics*[width=0.495\textwidth]{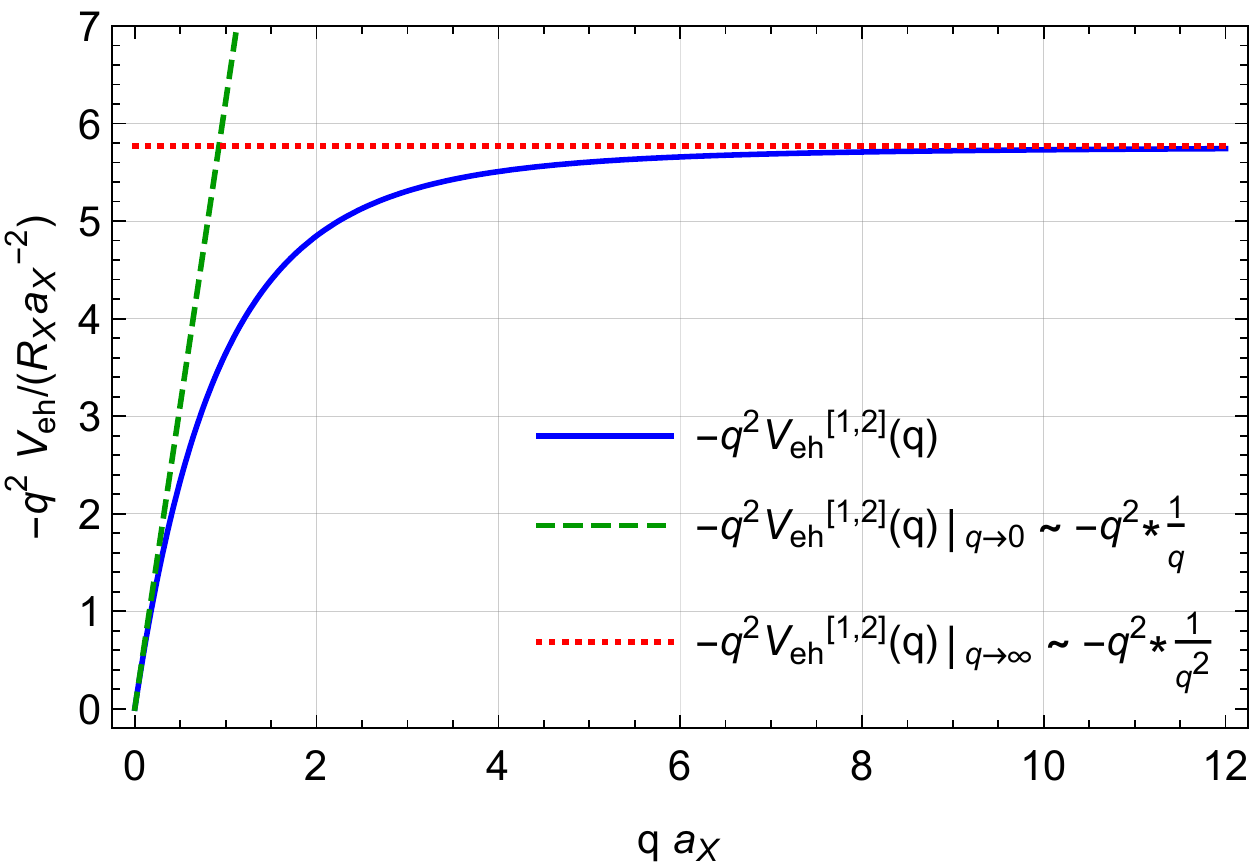}
\caption{Left panel: effective quasi-two-dimensional Coulomb potential (\ref{veh2}) between even states of electrons \textit{vs.} wave number $q$ ($a_{\rm X}$ is the (3d) excitonic Bohr radius and $R_{\rm X}$ the (3d) excitonic Rydberg energy), right panel: same quantity times square of the wave number $q$ compared to the 2d and 3d limiting cases.
}
\label{fig:veh}
\end{center}
\end{figure}

The effective Coulomb potential (\ref{veh2}) is shown in figure \ref{fig:veh} for various even electron bound states in the well. (Keep in mind that here single-particle bound states in the well are meant. They must not be confused with electron-hole (two-particle) bound states, i.e. excitons.) We denote them by the quantum number pairs $[e,e]$ with $e=1,2,3$. Remember that $Q=qd/2$ in  (\ref{veh2}).

In the right panel, the effective potential is compared to the limiting cases for two ($V_{\rm ee}^{\rm 2d}\propto 1/q$) and three dimensions ($V_{\rm ee}^{\rm 3d}\propto 1/q^2$).

\section{Derivation of the polarization function in the one-dimensional case}

In a 1d system, the polarization function is given by
\begin{eqnarray}\label{Pi1-1d}
\Pi_{aa}^{(1d)}(k,\omega)=\frac{1}{2\pi}n_a\Lambda_a\,\exp\left(-\frac{\hbar^2k^2}{8m_ak_{\rm B}T_a}\right)\,I
\end{eqnarray}
with
\begin{eqnarray}\label{I1a-1d-cart}
I&=&\int\limits_{-\infty}^{\infty}\mathrm{d}q\,\mathrm{e}^{-\beta aq^2}
\left(\mathrm{e}^{\beta akq}-\mathrm{e}^{-\beta akq}\right)
\left[
\frac{w-2akq}{(w-2ax)^2+\epsilon^2}
-\frac{\mathrm{i}\,\epsilon}{(w-2akq)^2+\epsilon^2}\right]\nonumber\\
&=&I_1+I_2
\,.
\end{eqnarray}
We introduce the abbreviations $x=akq$ and $u=\beta/(ak^2)$ making the integral easier readable,
\begin{eqnarray}\label{I1a-1d-cart-1}
I&=&\frac{1}{ak}\int\limits_{-\infty}^{\infty}\mathrm{d}x\,
\mathrm{e}^{-ux^2}
\,\left(\mathrm{e}^{\beta x}-\mathrm{e}^{-\beta x}\right)
\left[
\frac{w-2x}{(w-2x)^2+\epsilon^2}-
\frac{\mathrm{i}\,\epsilon}{(w-2x)^2+\epsilon^2}\right]\nonumber\\
&=&I_1+I_2
\,.
\end{eqnarray}

The second (imaginary) contribution $I_2$ can again be calculated straightforwardly. Performing the limit $\epsilon\to 0$ leads to
\begin{eqnarray}\label{I2-1d-cart}
I_2&=&-\frac{\mathrm{i}\pi}{ak}\int\limits_{-\infty}^{\infty}\mathrm{d}x\,
\mathrm{e}^{-ux^2}
\,\left(\mathrm{e}^{\beta x}-\mathrm{e}^{-\beta x}\right)
\delta(w-2x)\nonumber\\
&=&-\frac{\mathrm{i}\pi}{2ak}\,\exp\left(-\frac{\beta w^2}{4ak^2}\right)
\,\left[\exp\left(\frac{\beta w}{2}\right)-\exp\left(-\frac{\beta w}{2}\right)\right]
\,.
\end{eqnarray}

In the first (real) contribution to $I$ (\ref{I1a-1d-cart-1}) we proceed as before applying the integration trick (\ref{trick}),
\begin{eqnarray}\label{I11-1d-cart}
I_1&=&\frac{1}{ak}\int\limits_{-\infty}^{\infty}\mathrm{d}x\,
\mathrm{e}^{-ux^2}
\left(\mathrm{e}^{\beta x}-\mathrm{e}^{-\beta x}\right)(w-2x)
%\nonumber\\
%&&\hspace*{15ex}\times
\int\limits_0^{\infty}\mathrm{d}z\,\exp\left([(w-2x)^2+\epsilon^2]z\right)\nonumber\\
&=&\frac{1}{ak}\int\limits_0^{\infty}\mathrm{d}z\,\exp\left(-(w^2+\epsilon^2)z\right)
\int\limits_{-\infty}^{\infty}\mathrm{d}x\,\exp\left(-(u+4z)x^2\right)
\nonumber\\
&&\times
\left(\mathrm{e}^{(\beta+4z)x}-\mathrm{e}^{-(\beta-4z)x}\right)(w-2x)\nonumber\\
&=&\frac{1}{ak}\int\limits_0^{\infty}\mathrm{d}z\,\exp\left((w^2+\epsilon^2)z\right)\,I_1^{(xy)}(z)
\,.
\end{eqnarray}
The result of the $x$- and $y$-integrations is simple compared to the 2d case,
\begin{eqnarray}\label{I11-1d-cart-1}
\fl I_1^{(xy)}(z)&=&
\frac{\sqrt{\pi }}{(u+4 z)^{3/2}}\\
\fl &&\times\left\{(u w-\beta )\exp\left[\frac{(\beta +4 w z)^2}{4 (u+4 z)}\right]-(uw+\beta)\exp\left[\frac{(\beta -4 w z)^2}{4 (u+4 z)}\right]\right\}\,.\nonumber
\end{eqnarray}
We insert (\ref{I11-1d-cart-1}) into (\ref{I11-1d-cart}) and perform the limit $\epsilon\to 0$, yielding finally
\begin{eqnarray}\label{I11-1d-cart-3}
I_1&=&\frac{\sqrt{\pi}}{ak}\int\limits_0^{\infty}\mathrm{d}z\,\mathrm{e}^{-w^2z}\,
\frac{1}{(u+4 z)^{3/2}}\\
&&\times\left\{(u w-\beta )\exp\left[\frac{(\beta +4 w z)^2}{4 (u+4 z)}\right]-(uw+\beta)\exp\left[\frac{(\beta -4 w z)^2}{4 (u+4 z)}\right]\right\}\nonumber\\
&=&\frac{\sqrt{\pi }}{ak}\,\exp\left(\frac{\beta ^2}{4 u}\right) \left[F\left(\frac{u w-\beta }{2 \sqrt{u}}\right)-F\left(\frac{u w+\beta }{2 \sqrt{u}}\right)\right]\nonumber\\
&=&\frac{\sqrt{\pi }}{ak}\,\exp\left(\frac{\beta ak^2}{4}\right)
\left\{
F\left[\frac{\sqrt{\beta}}{2\sqrt{a}k}\left(w-ak^2\right)\right]
-F\left[\frac{\sqrt{\beta}}{2\sqrt{a}k}\left(w+ak^2\right)\right]
\right\}\nonumber
\end{eqnarray}
with Dawson's integral $F$.

Therefore, the final result for the sum of real and imaginary parts again can be expressed in terms of the Faddeeva function w,
\begin{eqnarray}\label{Ifinal-1d}
&&\mathrm{e}^{-\frac{\beta ak^2}{4}}I=\mathrm{e}^{-\frac{\beta ak^2}{4}}(I_1+I_2)=\nonumber\\
&&=-\frac{\pi}{2ak}
\Bigg(
\frac{2}{\sqrt{\pi}}\left\{
F\left[\frac{\sqrt{\beta}}{2\sqrt{a}k}\left(w+ak^2\right)\right]
-F\left[\frac{\sqrt{\beta}}{2\sqrt{a}k}\left(w-ak^2\right)\right]
\right\}\nonumber\\
&&\hspace*{10ex}+\mathrm{i}\left\{
\mathrm{exp}\left[-\frac{\beta}{4ak^2}\left(w-ak^2\right)^2\right]
-\mathrm{exp}\left[-\frac{\beta}{4ak^2}\left(w+ak^2\right)^2\right]
\right\}
\Bigg)\nonumber\\
&&=\frac{\mathrm{i}\pi}{2ak}
\Bigg\{
\mathrm{w}\left[\frac{\sqrt{\beta}}{2\sqrt{a}k}\left(w+ak^2\right)\right]
-\mathrm{w}\left[\frac{\sqrt{\beta}}{2\sqrt{a}k}\left(w-ak^2\right)\right]
\Bigg\}\,,
\end{eqnarray}
and the polarization function reads
\begin{eqnarray}\label{Pi2-1d}
\fl \Pi_{aa}^{(1d)}(k,\omega)
%=\nonumber\\
=\frac{1}{(2\pi)^2}n_a\Lambda_a\,\frac{\mathrm{i}\pi}{2ak}
\Bigg\{
\mathrm{w}\left[\frac{\sqrt{\beta}}{2\sqrt{a}k}\left(w+ak^2\right)\right]
-\mathrm{w}\left[\frac{\sqrt{\beta}}{2\sqrt{a}k}\left(w-ak^2\right)\right]
\Bigg\}\nonumber\\
\fl =\mathrm{i}\frac{\sqrt{\pi}}{2}n_a\sqrt{\frac{2m_a}{\hbar^2k^2}}\frac{1}{\sqrt{k_{\rm B}T_a}}\nonumber\\
\fl \times
\Bigg\{
\mathrm{w}\left[\frac{1}{2\sqrt{k_{\rm B}T_a}}\sqrt{\frac{2m_a}{\hbar^2k^2}}\left(\hbar\omega+\frac{\hbar^2k^2}{2m_a}\right)\right]
-\mathrm{w}\left[\frac{1}{2\sqrt{k_{\rm B}T_a}}\sqrt{\frac{2m_a}{\hbar^2k^2}}\left(\hbar\omega-\frac{\hbar^2k^2}{2m_a}\right)\right]
\Bigg\}\,.
%\nonumber\\
\end{eqnarray}

Comparing this with the 3d and 2d results (\ref{Pi2}) and (\ref{Pi2-2d}), we find that $\Pi_{aa}^{(1d)}(k,\omega)=\Pi_{aa}^{(2d)}(k,\omega)=\Pi_{aa}^{(3d)}(k,\omega)$.

\section{Derivation of the polarization function for linear carrier dispersion}\label{app:lindisp}

We consider the case of quasi-two-dimensional systems with linear carrier dispersions, $E(\mathbf{k})=\gamma k$, without an energy gap between ``valence'' and ``conduction'' band which resemble the Dirac cones of relativistic massless fermions. In such a system, interband transitions have to be taken into account, and the expression for the dielectric function (\ref{epsilon}) has to be modified into
\begin{equation}\label{epsilon-linear}
\varepsilon(k,\omega)=1-V_{ee}(k)\,\sum\limits_{a,b}\Pi_{ab}(k,\omega)
\end{equation}
where $a,b=\pm 1$ denote electrons in the upper ($+$) and lower ($-$) cone, respectively.
The polarization function reads
\begin{eqnarray}\label{Pilin1}
\fl \Pi_{ab}(k,\omega)&=&\frac{2}{(2\pi)^2}
%\nonumber\\
%&&\times
\int\mathrm{d}^2q\,\mathcal{F}_{ab}(\mathbf{k},\mathbf{q})
\frac{f_b\left(\left|\frac{\mathbf{k}}{2}-\mathbf{q}\right|\right)
-f_a\left(\left|\frac{\mathbf{k}}{2}+\mathbf{q}\right|\right)}
{w+E_b\left(\left|\frac{\mathbf{k}}{2}-\mathbf{q}\right|\right)-E_a\left(\left|\frac{\mathbf{k}}{2}+\mathbf{q}\right|\right)+\mathrm{i}\epsilon}\nonumber\\
\fl&=&
\frac{1}{2\pi^2}\left\{
\mathcal{P}\int\mathrm{d}^2q\,\mathcal{F}_{ab}(\mathbf{k},\mathbf{q})
\frac{f_b\left(\left|\frac{\mathbf{k}}{2}-\mathbf{q}\right|\right)
-f_a\left(\left|\frac{\mathbf{k}}{2}+\mathbf{q}\right|\right)}
{w+E_b\left(\left|\frac{\mathbf{k}}{2}-\mathbf{q}\right|\right)-E_a\left(\left|\frac{\mathbf{k}}{2}+\mathbf{q}\right|\right)}
\right.\nonumber\\
\fl&&
-\mathrm{i}\pi
\int\mathrm{d}^2q\,\mathcal{F}_{ab}(\mathbf{k},\mathbf{q})\,
\left[f_b\left(\left|\frac{\mathbf{k}}{2}-\mathbf{q}\right|\right)
-f_a\left(\left|\frac{\mathbf{k}}{2}+\mathbf{q}\right|\right)\right]\nonumber\\
\fl&&\times\left.\delta\left(w+E_b\left(\left|\frac{\mathbf{k}}{2}-\mathbf{q}\right|\right)-E_a\left(\left|\frac{\mathbf{k}}{2}+\mathbf{q}\right|\right)\right)
\right\}\,.
\end{eqnarray}
The band overlap factor $\mathcal{F}_{ab}(\mathbf{k},\mathbf{q})$ arises from degenerate bands like in graphene, i.e.,
\begin{eqnarray}\label{overlap}
\mathcal{F}_{ab}(\mathbf{k},\mathbf{q})=\left\{
\begin{array}{ll}
\frac{1}{2}(1+ab\cos\theta_{\mathbf{q},\mathbf{k}+\mathbf{q}}) & \quad\mbox{if band overlap is considered}\\
1 & \quad\mbox{otherwise}\,.
\end{array}
\right.
\end{eqnarray}
Obviously, the energy differences are given by
\begin{eqnarray}
E_{\pm}(k)-E_{\pm}(k')=\pm\gamma(k-k')\,,\quad
E_{\pm}(k)-E_{\mp}(k')=\pm\gamma(k+k')\,,
\end{eqnarray}
i.e., $E_b(k)-E_a(k')=b\gamma(k-abk')$. Concerning the occupation of the bands we consider the case of weak excitation above the ground state with $\mu=0$, i.e., a small number of electrons is excited from the lower into the upper cone, e.g., by a laser. Then we have in the upper cone an electron distribution which can be well approximated by a Boltzmann distribution, $f_{+}(k)\approx \frac{n\Lambda^2}{2}\,\mathrm{e}^{-\beta\gamma k}$, and the distribution of the electrons in the lower cone is given by $f_{-}(k)=1-f_{+}(k)\approx 1-\frac{n\Lambda^2}{2}\,\mathrm{e}^{-\beta\gamma k}$. 

First we look at the imaginary part of $\Pi_{ab}$ [last summand of (\ref{Pilin1})]. After inserting the distribution function and energy differences, the expression does not look much more complicated than in the case of parabolic dispersion [cf.\ (\ref{Pi1})], however,  the absolute values of wave number differences turn into square roots, and the expression cannot be integrated straightforwardly. Instead, we first introduce an additional auxiliary integration by substituting the variables
\begin{eqnarray}
\gamma\left(\frac{\mathbf{k}}{2}-\mathbf{q}\right)=\mathbf{x}\,,\quad
\gamma\left(\frac{\mathbf{k}}{2}+\mathbf{q}\right)=\mathbf{y}\,,\quad\mbox{i.e.,}\quad
\mathbf{x}+\mathbf{y}=\gamma\mathbf{k}\,.
\end{eqnarray}
Im $\Pi_{ab}$ then reads
\begin{eqnarray}\label{ImPilin1}
\mathrm{Im}\,\Pi_{ab}(k,\omega)=&-\frac{1}{2\pi\gamma^2}
\int\mathrm{d}^2x\int\mathrm{d}^2y\,
\left[f_b(x)-f_a(y)\right]
\mathcal{F}_{ab}(\mathbf{x},\mathbf{y})
\nonumber\\
&\times\delta[w+b(x-aby)]\,\delta(\mathbf{y}-(\gamma\mathbf{k}-\mathbf{x}))
\end{eqnarray}
with
\begin{eqnarray}
\fl\mathcal{F}_{ab}(\mathbf{x},\mathbf{y})=\mathcal{F}_{ab}(x,y,\cos\varphi)=\left\{
\begin{array}{l}
\frac{1}{2}(1+ab\cos\theta_{-\mathbf{x},\gamma\mathbf{k}-\mathbf{x}})
=\frac{1}{2}\!\left[1+ab\frac{\gamma k\,\cos\,\varphi}{(\gamma^2k^2+x^2-2\gamma kx\,\cos\varphi)^{1/2}}\right]\\
1
\end{array}
\right.\nonumber\\
\end{eqnarray}
[see (\ref{overlap})], where $\varphi=\varphi_x=\measuredangle(\mathbf{k},\mathbf{x})$.

The expression (\ref{ImPilin1}) is easier to handle, however, the last delta function has to be considered very carefully. We introduce polar coordinates (the abscissa for the $x$-integration is chosen in $k$-direction):
\begin{eqnarray}\label{ImPilin2}
\fl\mathrm{Im}\,\Pi_{ab}(k,\omega)&=&-\frac{1}{2\pi\gamma^2}
\int\limits_0^{\infty}\mathrm{d}x\,x\int\limits_0^{\infty}\mathrm{d}y\,y
\int\limits_0^{2\pi}\mathrm{d}\varphi_x\int\limits_0^{2\pi}\mathrm{d}\varphi_y
\left[f_b(x)-f_a(y)\right]
\mathcal{F}_{ab}(x,y,\cos\varphi_x)\nonumber\\
&&\times\delta[w+b(x-aby)]\,
\frac{1}{y}\delta(y-|\gamma\mathbf{k}-\mathbf{x}|)\delta(\varphi_y-\varphi_x)\nonumber\\
&=&-\frac{1}{2\pi\gamma^2}
\int\limits_0^{\infty}\mathrm{d}x\,x\int\limits_0^{\infty}\mathrm{d}y\int\limits_0^{2\pi}\mathrm{d}\varphi
\left[f_b(x)-f_a(y)\right]
\mathcal{F}_{ab}(x,y,\cos\varphi)\nonumber\\
&&\times\delta[w+b(x-aby)]
\delta\left[y-\left(\gamma^2k^2+x^2-2\gamma k x\,\cos\,\varphi\right)^{1/2}\right].
\end{eqnarray}
To perform the $\varphi$-integration we use the last delta function,
\begin{eqnarray}\label{deltaphi}
\delta\!\left[y-\left(\gamma^2k^2+x^2-2\gamma k x\,\cos\,\varphi\right)^{1/2}\right]
%=\delta\left(f(\varphi)\right)\nonumber\\
=\sum\limits_{i=1}^2\frac{1}{|f'(\varphi)|_{\varphi=\varphi_i}}\delta(\varphi-\varphi_i)
%=\frac{1}{|f'(\varphi)|_{\varphi=\varphi_0}}\delta(\varphi-\varphi_0)
\end{eqnarray}
with
\begin{eqnarray}\label{deltaphi1}
\cos\,\varphi_{1/2}=\frac{1}{2\gamma kx}(\gamma^2k^2+x^2-y^2)\,,\nonumber\\
|f'(\varphi)|=\frac{\gamma kx\,|\sin\varphi|}{\left(\gamma^2k^2+x^2-2\gamma k x\,\cos\,\varphi\right)^{1/2}}\,,\nonumber\\
|f'(\varphi)|_{\varphi=\varphi_{1/2}}
%=\frac{\sqrt{4\gamma^2k^2x^2-(\gamma^2k^2-w^2-2wx)^2}}{2(w+x)}
=\frac{\left\{\left[(x+y)^2-\gamma^2k^2\right]
\left[\gamma^2k^2-(x-y)^2\right]\right\}^{1/2}}{2y}.
\end{eqnarray}
From the conditions for real solutions (i) $|\cos\,\varphi_{1/2}|\le 1$ and (ii) positive radicand in the last line of (\ref{deltaphi1}), we find the restriction
$|\gamma k-x|\le y\le\gamma k+x$.

Inserting now (\ref{deltaphi}) and (\ref{deltaphi1}) into (\ref{ImPilin2}), we obtain
\begin{eqnarray}\label{ImPilin3}
\mathrm{Im}\,\Pi_{ab}(k,\omega)&=&-\frac{2}{\pi\gamma^2}
\int\limits_0^{\infty}\mathrm{d}x\int\limits_{|\gamma k-x|}^{\gamma k+x}\mathrm{d}y\,
\left[f_b(x)-f_a(y)\right]
\delta[w+b(x-aby)]\nonumber\\
&&\times
\frac{x\,y\,\mathcal{F}_{ab}(x,y)}{\left\{\left[(x+y)^2-\gamma^2k^2\right]
\left[\gamma^2k^2-(x-y)^2\right]\right\}^{1/2}}
\end{eqnarray}
with
\begin{eqnarray}
\mathcal{F}_{ab}(x,y)=\left\{
\begin{array}{l}
\frac{1}{2}\left[1+\frac{ab}{2xy}(\gamma^2k^2-x^2-y^2)\right]\\
1\,.
\end{array}
\right.
\end{eqnarray}
Introducing dimensionless sum and difference variables, $z=(x+y)/(\gamma k)$ and $t=(y-x)/(\gamma k)$, respectively, and abbreviating $u=\beta\gamma k/2$ and $v=w/(\gamma k)$, (\ref{ImPilin3}) turns into
\begin{eqnarray}\label{ImPilin4}
\mathrm{Im}\,\Pi_{ab}(k,\omega)&=&-\frac{1}{\pi\gamma^2}
\frac{\gamma k}{4}
\int\limits_1^{\infty}\mathrm{d}z\int\limits_{-1}^{1}\mathrm{d}t\,
\left[f_b\left(\frac{z-t}{2}\right)-f_a\left(\frac{z+t}{2}\right)\right]
\nonumber\\
&&\times\delta\left(v-b\left\{\begin{array}{c}
                        \!t\!\\\!z\!
                       \end{array}
\right\}\right)
\frac{(z^2-t^2)\,\mathcal{F}_{ab}(z,t)}{\left[(z^2-1)(1-t^2)\right]^{1/2}}
\end{eqnarray}
with
\begin{eqnarray}
\mathcal{F}_{ab}(z,t)=\left\{
\begin{array}{ll}
\frac{z^2-1}{z^2-t^2} & ,\quad ab=+1, \mbox{intraband transitions}\\
\frac{1-t^2}{z^2-t^2} & ,\quad ab=-1, \mbox{interband transitions}\\
1 & ,\quad\mbox{no band overlap considered}\,,
\end{array}
\right.
\end{eqnarray}
so that the last fraction in (\ref{ImPilin4}) reads
\begin{eqnarray}\label{fraction}
\fl\frac{(z^2-t^2)\,\mathcal{F}_{ab}(z,t)}{\left[(z^2-1)(1-t^2)\right]^{1/2}}=\left\{
\begin{array}{ll}
\left[\frac{z^2-1}{1-t^2}\right]^{1/2} & ,\quad ab=+1, \mbox{intraband}\\
\left[\frac{1-t^2}{z^2-1}\right]^{1/2} & ,\quad ab=-1, \mbox{interband}\\
\left[\frac{z^2-1}{1-t^2}\right]^{1/2}+\left[\frac{1-t^2}{z^2-1}\right]^{1/2} & ,\quad\mbox{no band overlap}\,.
\end{array}
\right.
\end{eqnarray}
In the following, we consider the case with band overlap only.
Inserting (\ref{fraction}) and the difference of distribution functions into (\ref{ImPilin4}), we obtain for the intraband contributions to Im $\Pi$
\begin{eqnarray}\label{ImPilin5}
\fl\mathrm{Im}\,\Pi_{++}(k,\omega)&=&-\frac{n\Lambda^2}{\pi\gamma^2}
\frac{\gamma k}{8}
\int\limits_1^{\infty}\mathrm{d}z\,\mathrm{e}^{-uz}\sqrt{z^2-1}\,\int\limits_{-1}^{1}\mathrm{d}t\,
\frac{\mathrm{e}^{ut}-\mathrm{e}^{-ut}}{\sqrt{1-t^2}}
\delta(v-t)\nonumber\\
&=&-\frac{n\Lambda^2}{\pi\gamma^2}
\frac{\gamma k}{8}\,\frac{K_1(u)}{u}\,\frac{\mathrm{e}^{uv}-\mathrm{e}^{-uv}}{\sqrt{1-v^2}}\,
\Theta(1-v)\Theta(1+v)\nonumber\\
&=&\mathrm{Im}\,\Pi_{--}(k,\omega)\,,
\end{eqnarray}
where $K_0$ and $K_1$ denote modified Bessel functions of the second kind, also referred to as MacDonald functions or modified Hankel functions (i.e., Hankel functions with imaginary arguments, $K_\nu(x)=\frac{\pi}{2}\mathrm{i}^{\nu+1}H_\nu^{(1)}(\mathrm{i}x)$ \cite{abramowitz}). For the interband contributions we get
\begin{eqnarray}\label{ImPilin6}
\fl\mathrm{Im}\,\Pi_{+-}(k,\omega)&=&\frac{n\Lambda^2}{\pi\gamma^2}
\frac{\gamma k}{8}
\int\limits_1^{\infty}\mathrm{d}z\,\frac{\mathrm{e}^{-uz}}{\sqrt{z^2-1}}\,\delta(v-z)\,\int\limits_{-1}^{1}\mathrm{d}t\,
\left[\mathrm{e}^{ut}+\mathrm{e}^{-ut}\right]\sqrt{1-t^2}
\nonumber\\
&&-\frac{1}{\pi\gamma^2}
\frac{\gamma k}{4}
\int\limits_1^{\infty}\mathrm{d}z\,\frac{1}{\sqrt{z^2-1}}\,\delta(v-z)\,\int\limits_{-1}^{1}\mathrm{d}t\,
\sqrt{1-t^2}\nonumber\\
&=&\frac{1}{\gamma^2}
\frac{\gamma k}{4}\left\{n\Lambda^2\frac{\mathrm{e}^{-uv}}{\sqrt{v^2-1}}\,\frac{I_1(u)}{u}-\frac{1}{2}\frac{1}{\sqrt{v^2-1}}\right\}
\Theta(v-1)\,,\\
\label{ImPilin7}
\fl\mathrm{Im}\,\Pi_{-+}(k,\omega)&=&-\frac{n\Lambda^2}{\pi\gamma^2}
\frac{\gamma k}{8}
\int\limits_1^{\infty}\mathrm{d}z\,\frac{\mathrm{e}^{-uz}}{\sqrt{z^2-1}}\,\delta(v+z)\,\int\limits_{-1}^{1}\mathrm{d}t\,
\left[\mathrm{e}^{ut}+\mathrm{e}^{-ut}\right]\sqrt{1-t^2}
\nonumber\\
&&+\frac{1}{\pi\gamma^2}
\frac{\gamma k}{4}
\int\limits_1^{\infty}\mathrm{d}z\,\frac{1}{\sqrt{z^2-1}}\,\delta(v+z)\,\int\limits_{-1}^{1}\mathrm{d}t\,
\sqrt{1-t^2}\nonumber\\
&=&-\frac{1}{\gamma^2}
\frac{\gamma k}{4}\left\{n\Lambda^2\frac{\mathrm{e}^{-uv}}{\sqrt{v^2-1}}\,\frac{I_1(u)}{u}-\frac{1}{2}\frac{1}{\sqrt{v^2-1}}\right\}
\Theta(-v-1)\,,
\end{eqnarray}
where $I_n$ denotes the modified Bessel function of first kind (i.e., Bessel functions with imaginary arguments, $I_\nu(x)=\mathrm{i}^{-\nu}J_\nu(\mathrm{i}x)$) \cite{abramowitz}.

Looking back at (\ref{fraction}), in the case without band overlap, there would occur the double number of terms in the contributions to Im $\Pi$ considered above. However, the additional terms lead to divergencies which casts the physical sense of this case into doubt.

% $\mathrm{Im}\,\Pi_{aa}$ then finally reads
% \begin{eqnarray}\label{ImPilin5app}
% \mathrm{Im}\,\Pi_{aa}(k,\omega)
% =&-\frac{k}{8\pi\gamma}\,n_a\Lambda_a^2
% \left[\exp\left(\frac{\beta\hbar\omega}{2}\right)-\exp\left(-\frac{\beta\hbar\omega}{2}\right)\right]\\
% %\nonumber\\
% &\times
% \left\{
% \frac{2}{\beta\gamma k}
% \frac{K_1\left(\frac{\beta\gamma k}{2}\right)}{\sqrt{1-\left(\frac{\hbar\omega}{\gamma k}\right)^2}}
% +K_0\left(\frac{\beta\gamma k}{2}\right)\sqrt{1-\left(\frac{\hbar\omega}{\gamma k}\right)^2}
% \right\}.\nonumber
% \end{eqnarray}

The real part of the polarization function can be obtained either via Kramers--Kronig transformation of the imaginary part or, equivalently, directly from (\ref{Pilin1}). Since the calculation runs analogously, we have only to replace the remaining delta functions in (\ref{ImPilin4}) or (\ref{ImPilin5})--(\ref{ImPilin7}), respectively, by the corresponding denominator, e.g., $\delta(v-t)\to\frac{1}{v-t}$ etc. Obviously, then one integral remains in each case,
\begin{eqnarray}\label{RePilin5}
\mathrm{Re}\,\Pi_{++}(k,\omega)&=&\frac{n\Lambda^2}{\pi^2\gamma^2}
\frac{\gamma k}{8}
\frac{K_1(u)}{u}\,\,
\mathcal{P}\int\limits_{-1}^{1}\mathrm{d}t\,
\frac{\mathrm{e}^{ut}-\mathrm{e}^{-ut}}{\sqrt{1-t^2}}
\frac{1}{v-t}\nonumber\\
&=&\mathrm{Re}\,\Pi_{--}(k,\omega)\,,\\
\label{RePilin6}
\mathrm{Re}\,\Pi_{+-}(k,\omega)&=&
-\frac{1}{\pi\gamma^2}
\frac{\gamma k}{4}\left\{n\Lambda^2\,\frac{I_1(u)}{u}\,\mathcal{P}\int\limits_1^{\infty}\mathrm{d}z\,\frac{\mathrm{e}^{-uz}}{\sqrt{z^2-1}}\,\frac{1}{v-z}\right.\nonumber\\
&&\left.-\frac{1}{2}\,\mathcal{P}\int\limits_1^{\infty}\mathrm{d}z\,\frac{1}{\sqrt{z^2-1}}\,\frac{1}{v-z}\right\}\,,\nonumber\\\\
\label{RePilin7}
\mathrm{Re}\,\Pi_{-+}(k,\omega)&=&
\frac{1}{\pi\gamma^2}
\frac{\gamma k}{4}\left\{n\Lambda^2\,\frac{I_1(u)}{u}\,\mathcal{P}\int\limits_1^{\infty}\mathrm{d}z\,\frac{\mathrm{e}^{-uz}}{\sqrt{z^2-1}}\,\frac{1}{v+z}\right.\nonumber\\
&&\left.-\frac{1}{2}\,\mathcal{P}\int\limits_1^{\infty}\mathrm{d}z\,\frac{1}{\sqrt{z^2-1}}\,\frac{1}{v+z}\right\}\,.\nonumber\\
\end{eqnarray}
Combining intraband and interband contributions, respectively, we obtain
\begin{eqnarray}\label{RePilin8}
\mathrm{Re}\,\Pi_{\rm intra}(k,\omega)&=&
\mathrm{Re}\,\Pi_{++}(k,\omega)+\mathrm{Re}\,\Pi_{--}(k,\omega)\\
&=&-\frac{n\Lambda^2}{\pi^2\gamma^2}
\frac{\gamma k}{4}
\frac{K_1(u)}{u}\,\,
\mathcal{P}\int\limits_{-1}^{1}\mathrm{d}t\,
\frac{\mathrm{e}^{ut}-\mathrm{e}^{-ut}}{\sqrt{1-t^2}}
\frac{1}{t-v}\nonumber\\
&=&-\frac{n\Lambda^2}{\pi^2\gamma^2}
\frac{\gamma k}{4}
\frac{K_1(u)}{u}\,\,\mathcal{R}(u,v)\,,\nonumber\\
\label{RePilin9}
\mathrm{Re}\,\Pi_{\rm inter}(k,\omega)&=&
\mathrm{Re}\,\Pi_{+-}(k,\omega)+\mathrm{Re}\,\Pi_{-+}(k,\omega)\\
&=&-\frac{1}{\pi\gamma^2}
\frac{\gamma k}{4}\left\{n\Lambda^2\,\frac{I_1(u)}{u}\,\mathcal{P}\int\limits_1^{\infty}\mathrm{d}z\,\frac{\mathrm{e}^{-uz}}{\sqrt{z^2-1}}\left[\frac{1}{v-z}-\frac{1}{v+z}\right]\right.\nonumber\\
&&\left.-\frac{1}{2}\mathcal{P}\int\limits_1^{\infty}\mathrm{d}z\,\frac{1}{\sqrt{z^2-1}}\left[\frac{1}{v-z}-\frac{1}{v+z}\right]\right\}\nonumber\\
&=&-\frac{1}{\pi\gamma^2}
\frac{\gamma k}{4}\left\{n\Lambda^2\,\frac{I_1(u)}{u}\,\mathcal{S}(u,v)
+\frac{\pi}{2}\,\frac{1}{\sqrt{1-v^2}}\,\Theta(1-v)
\right\}\,.\nonumber
\end{eqnarray}

The remaining integrals $\mathcal{R}(u,v)$ and $\mathcal{S}(u,v)$ in (\ref{RePilin8}) and (\ref{RePilin9}) are very interesting mathematical objects. To further analyze them, we convert $\mathcal{S}(u,v)$ into a similar shape as $\mathcal{R}(u,v)$ by substituting $t=1/z$,
\begin{eqnarray}
\mathcal{S}(u,v)=\frac{1}{v}\,\mathcal{P}\int\limits_{-1}^{1}\mathrm{d}t\,
\frac{\mathrm{e}^{-\frac{u}{|t|}}}{\sqrt{1-t^2}}\,\frac{1}{t-\frac{1}{v}}
\end{eqnarray}
Resistant against all efforts to solve them analytically, $\mathcal{R}(u,v)$ and $\mathcal{S}(u,v)$ turn out to be equivalent to a certain kind of integral transformation between Chebyshev series of first and second kind. We use the following relation \cite{mason}:

%\begin{enumerate}
%\item 
\begin{eqnarray}\label{chebtrafo1}
\mbox{If}\quad f(y)\sim\sum\limits_{n=1}^{\infty}a_nT_n(y)\quad\mbox{and}\quad g(x)\sim\pi\sum\limits_{n=1}^{\infty}a_nU_{n-1}(x)\quad\mbox{then}\nonumber\\
g(x)=\mathcal{P}\int\limits_{-1}^1\mathrm{d}y\,
\frac{f(y)}{\sqrt{1-y^2}(y-x)}\,,
\end{eqnarray}
% \item If $f(y)\sim\sum_{n=1}^{\infty}b_nU_{n-1}(y)$ and $g(x)\sim\pi\sum_{n=1}^{\infty}b_nT_n(x)$ then
% \begin{eqnarray}\label{chebtrafo2}
% g(x)=\mathcal{P}\int\limits_{-1}^1\mathrm{d}y\,
% \frac{\sqrt{1-y^2}f(y)}{y-x}\,.
% \end{eqnarray}
% \end{enumerate}
$T_n$ and $U_n$ are the Chebyshev polynomials of first and second kind, respectively \cite{abramowitz}. Note that (\ref{chebtrafo1}) holds only for $|x|\le 1$.

Most importantly, the expansion coefficients of the series are the same on both sides. Therefore, if one knows the Chebyshev series (of first kind) of the numerator function, the series (of second kind) of the sought function is known, too.

In $\mathcal{R}(u,v)$, the numerator function is just the hyperbolic sine function. The coefficients of the Chebyshev series of first kind of $f(z,u)=\sinh\,uz$ are given by \cite{mason}
\begin{eqnarray}
a_n^{(\mathcal{R})}(u)=\frac{2}{\pi}\int\limits_{-1}^1\mathrm{d}z\,\sinh\,uz\,\frac{T_n(z)}{\sqrt{1-z^2}}\,
=\left\{
\begin{array}{ll}
0 & \mbox{if}\quad n=2k\\
2I_n(u) & \mbox{if}\quad n=2k+1\,,
\end{array}
\right.
\end{eqnarray}
where $I_n$ again denotes the modified Bessel function of first kind \cite{abramowitz}.

The relation (\ref{chebtrafo1}) has to be modified for $|x|>1$. In that case, $g(x)$ has to be complemented according to
\begin{eqnarray}\label{chebtrafo2}
g(x)\sim\pi\sum\limits_{n=1}^{\infty}a_n\left[U_{n-1}(x)-\frac{T_n(x)}{\sqrt{x^2-1}}\right]\,.
\end{eqnarray}
Therefore, one obtains for $\mathcal{R}(u,v)$ with (\ref{chebtrafo1})--(\ref{chebtrafo2}):
\begin{eqnarray}\label{R}
\mathcal{R}(u,v)&=&4\pi\sum\limits_{k=0}^{\infty}I_{2k+1}(u)\Bigg\{U_{2k}(v)\Theta(1-v)\Theta(1+v)\nonumber\\
&&\left.+\left[U_{2k}(v)-\frac{T_{2k+1}(v)}{\sqrt{v^2-1}}\right]\Theta(v-1)\Theta(-v-1)\right\}\,.
\end{eqnarray}

In $\mathcal{S}(u,v)$, the numerator function is $\mathrm{e}^{-\frac{u}{|t|}}$. The coefficients of the Chebyshev series of first kind have to be determined from
\begin{eqnarray}
a_n^{(\mathcal{S})}(u)=\frac{2}{\pi}\int\limits_{-1}^1\mathrm{d}z\,\mathrm{e}^{-\frac{u}{|z|}}\,\frac{T_n(z)}{\sqrt{1-z^2}}\,,
\end{eqnarray}
and $\mathcal{S}(u,v)$ is then given by
\begin{eqnarray}\label{S}
\mathcal{S}(u,v)&=&\frac{\pi}{v}\sum\limits_{k=1}^{\infty}a_k^{(\mathcal{S})}(u)\Bigg\{U_{k-1}(1/v)\Theta(1-1/v)\Theta(1+1/v)\nonumber\\
&&\left.+\left[U_{k-1}(1/v)-\frac{T_{k}(1/v)}{\sqrt{1/v^2-1}}\right]\Theta(1/v-1)\Theta(-1/v-1)\right\}\,.
\end{eqnarray}

Summarizing all contributions to imaginary and real parts of the polarization function of a graphene-like two-dimensional semiconductor structure, we have
\begin{eqnarray}\label{ImPilinfinal}
\fl\mathrm{Im}\,\Pi(k,\omega)&=&
\sum\limits_{a,b}\mathrm{Im}\,\Pi_{ab}(k,\omega)\nonumber\\
% &=&\frac{1}{\gamma^2}
% \frac{\gamma k}{8}\left\{\frac{n\Lambda^2}{\pi}
% \,\frac{\mathrm{e}^{-uv}-\mathrm{e}^{uv}}{\sqrt{1-v^2}}\,\frac{K_1(u)}{u}\,
% \Theta(1-v)\Theta(1+v)\right.\nonumber\\
% &&\left.+\left[n\Lambda^2\frac{\mathrm{e}^{-uv}}{\sqrt{v^2-1}}\,\frac{I_1(u)}{u}-\frac{1}{2}\frac{1}{\sqrt{v^2-1}}\right]
% \left[\Theta(v-1)-\Theta(-v-1)\right]
% \right\}\nonumber\\
\fl&=&\frac{1}{\gamma^2}
\frac{\gamma k}{4}\left\{\frac{n\Lambda^2}{\pi}
\,\frac{\mathrm{e}^{-\frac{\beta\hbar\omega}{2}}-\mathrm{e}^{\frac{\beta\hbar\omega}{2}}}{\sqrt{1-\left(\frac{\hbar\omega}{\gamma k}\right)^2}}\,\frac{K_1\left(\frac{\beta\gamma k}{2}\right)}{\beta\gamma k/2}\,
\Theta\left(1-\frac{\hbar\omega}{\gamma k}\right)\Theta\left(1+\frac{\hbar\omega}{\gamma k}\right)\right.\nonumber\\
\fl&&+\left[n\Lambda^2\frac{\mathrm{e}^{-\frac{\beta\hbar\omega}{2}}}{\sqrt{\left(\frac{\hbar\omega}{\gamma k}\right)^2-1}}\,\frac{I_1\left(\frac{\beta\gamma k}{2}\right)}{\beta\gamma k/2}-\frac{1}{2}\frac{1}{\sqrt{\left(\frac{\hbar\omega}{\gamma k}\right)^2-1}}\right]\nonumber\\
\fl&&\left.\times
\left[\Theta\left(\frac{\hbar\omega}{\gamma k}-1\right)-\Theta\left(-\frac{\hbar\omega}{\gamma k}-1\right)\right]
\right\}\,,
\end{eqnarray}

\begin{eqnarray}\label{RePilinfinal}
\fl\mathrm{Re}\,\Pi(k,\omega)&=&
\sum\limits_{a,b}\mathrm{Re}\,\Pi_{ab}(k,\omega)\\
\fl&=&-\frac{1}{\gamma^2}
\frac{\gamma k}{4}\left\{\frac{n\Lambda^2}{\pi^2}\,
\frac{K_1\left(\frac{\beta\gamma k}{2}\right)}{\beta\gamma k/2}\,\mathcal{R}\left(\frac{\beta\gamma k}{2},\frac{\hbar\omega}{\gamma k}\right)
+\frac{n\Lambda^2}{4\pi}\,\frac{I_1\left(\frac{\beta\gamma k}{2}\right)}{\beta\gamma k/2}\,\frac{\gamma k}{\hbar\omega}\,\mathcal{S}\left(\frac{\beta\gamma k}{2},\frac{\gamma k}{\hbar\omega}\right)\right.\nonumber\\
\fl&&\left.
% \nonumber\\
% \fl&&\left.
+\frac{1}{2\sqrt{1-\left(\frac{\hbar\omega}{\gamma k}\right)^2}}\,\Theta\left(1-\frac{\hbar\omega}{\gamma k}\right)\Theta\left(1+\frac{\hbar\omega}{\gamma k}\right)
\right\}\,,\nonumber
\end{eqnarray}
where $\mathcal{R}(u,v)$ and $\mathcal{S}(u,v)$ are given by (\ref{R}) and (\ref{S}).

% Both real and imaginary parts are nonzero only for $\hbar\omega\le\gamma k$, i.e., inside a cone spanned by the single-particle dispersion. This phenomenon resembles the similar Dirac-cone physics known from graphene and topological quantum matter, see, e.g. \cite{HD07,kotov2012,wunsch2006}.

\begin{figure}[h]%
\begin{center}
\includegraphics*[width=0.49\textwidth]{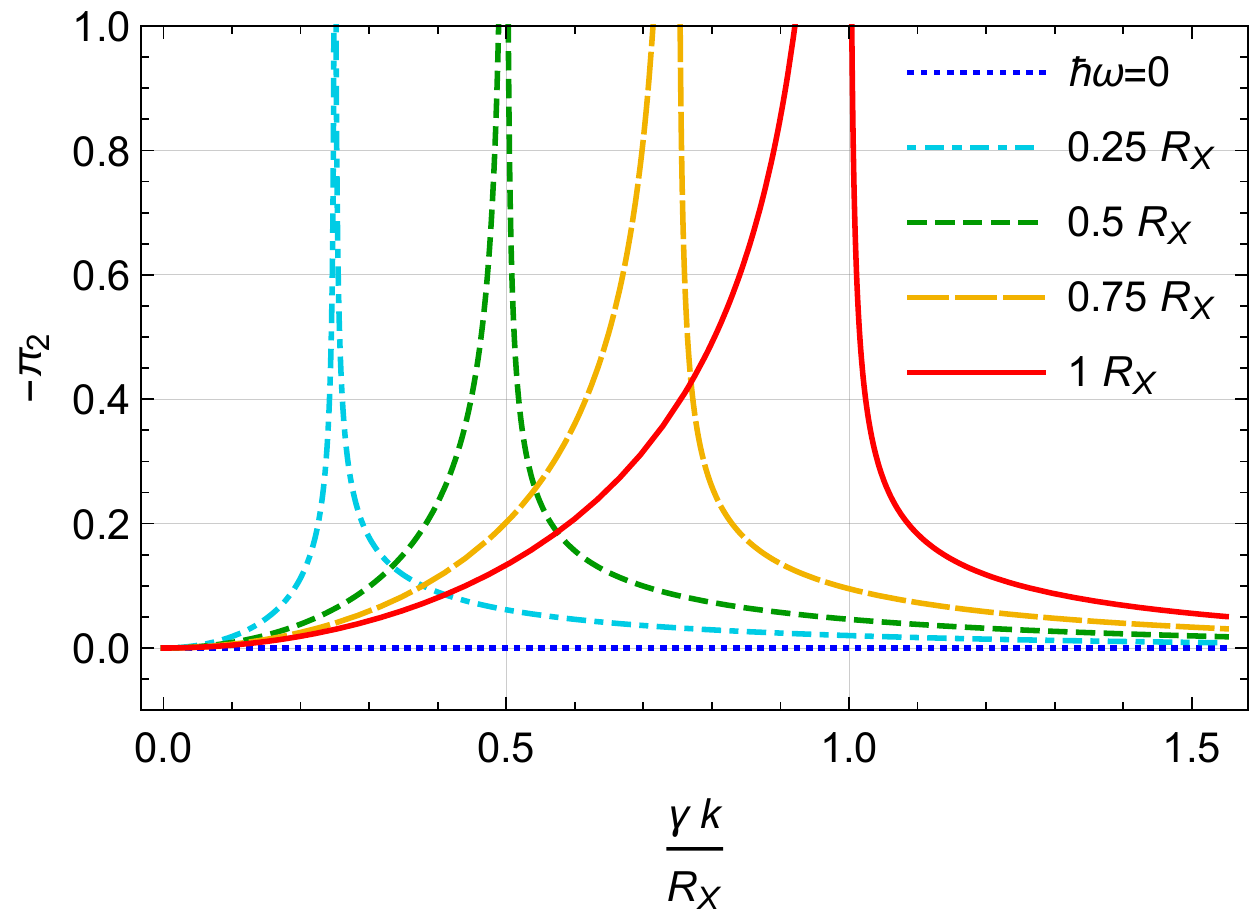}\hspace*{0.1cm}
\includegraphics*[width=0.49\textwidth]{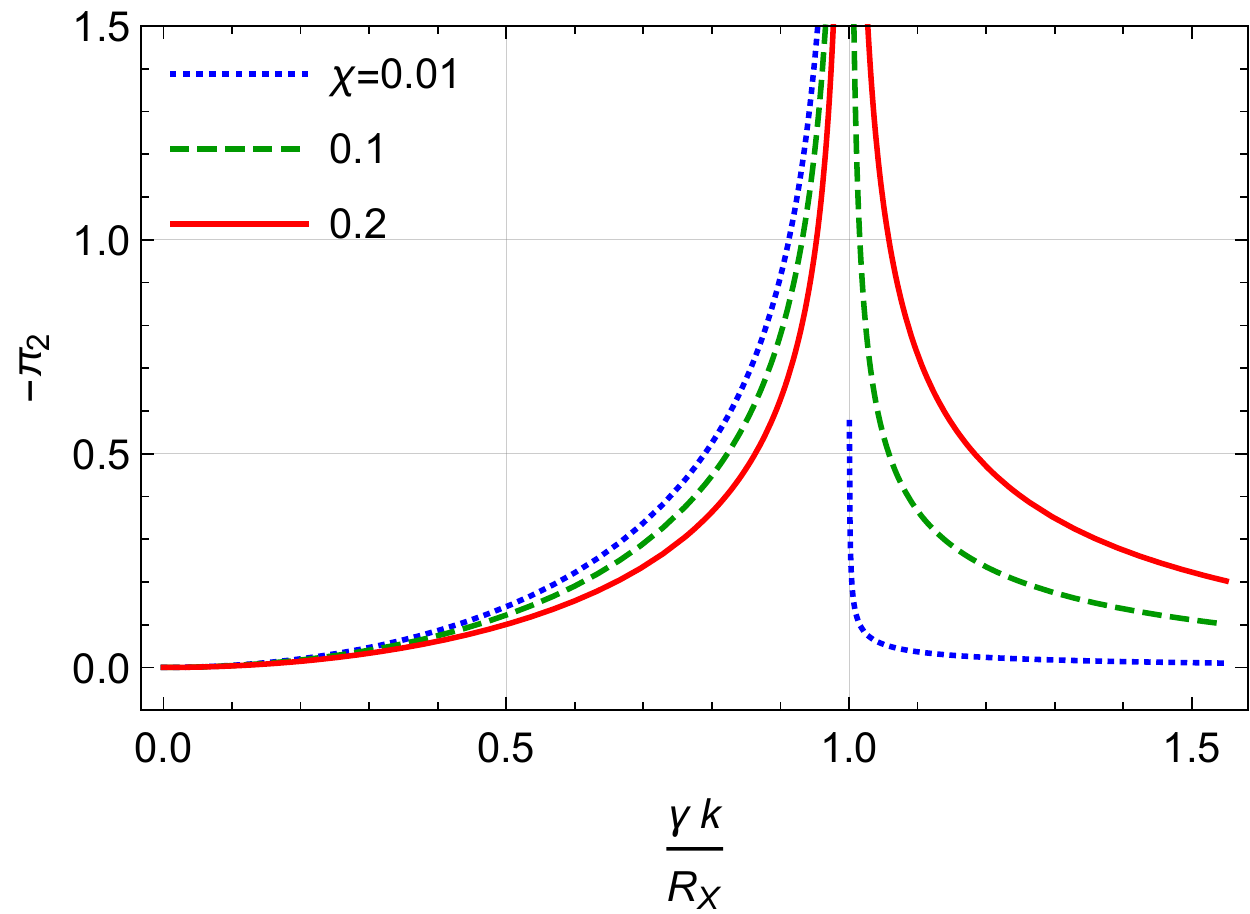}\\
\includegraphics*[width=0.49\textwidth]{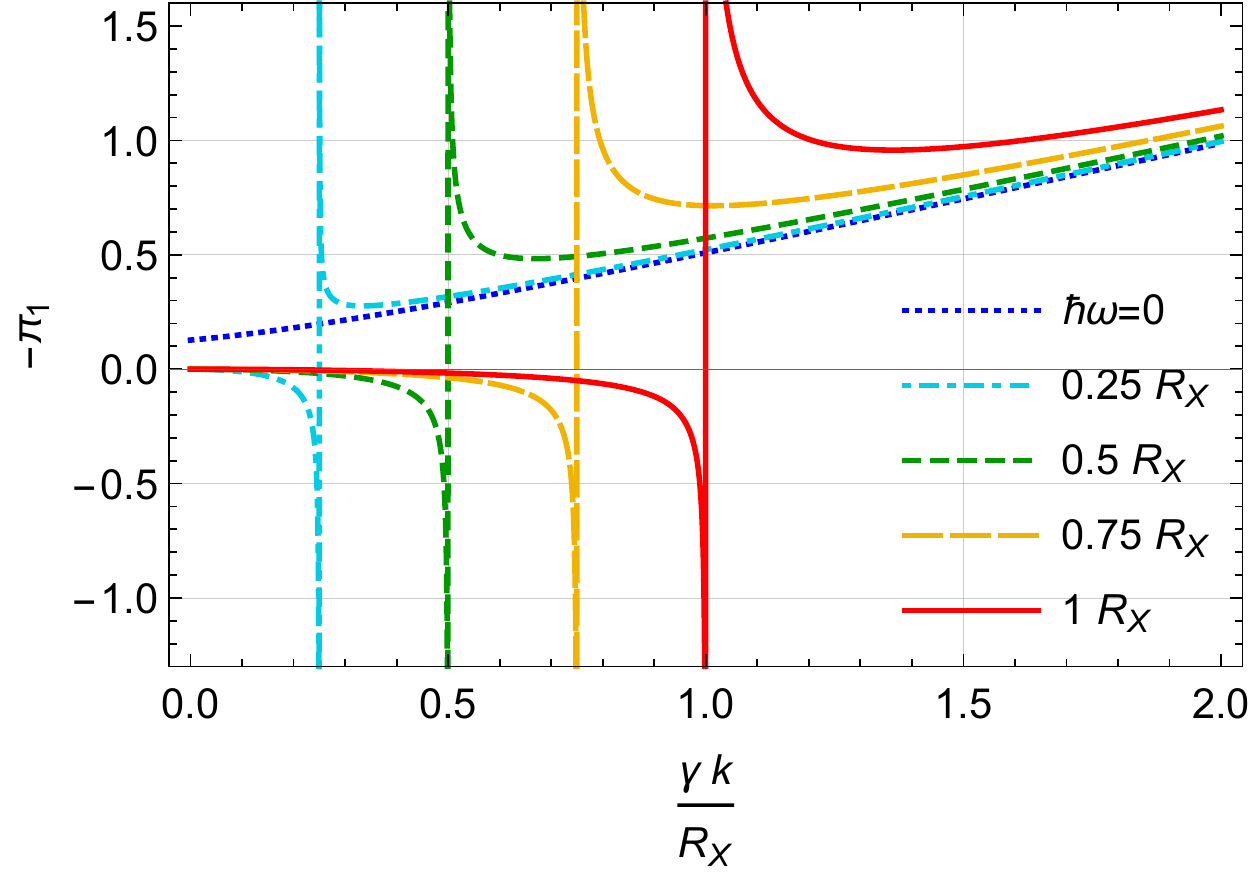}\hspace*{0.1cm}
\includegraphics*[width=0.49\textwidth]{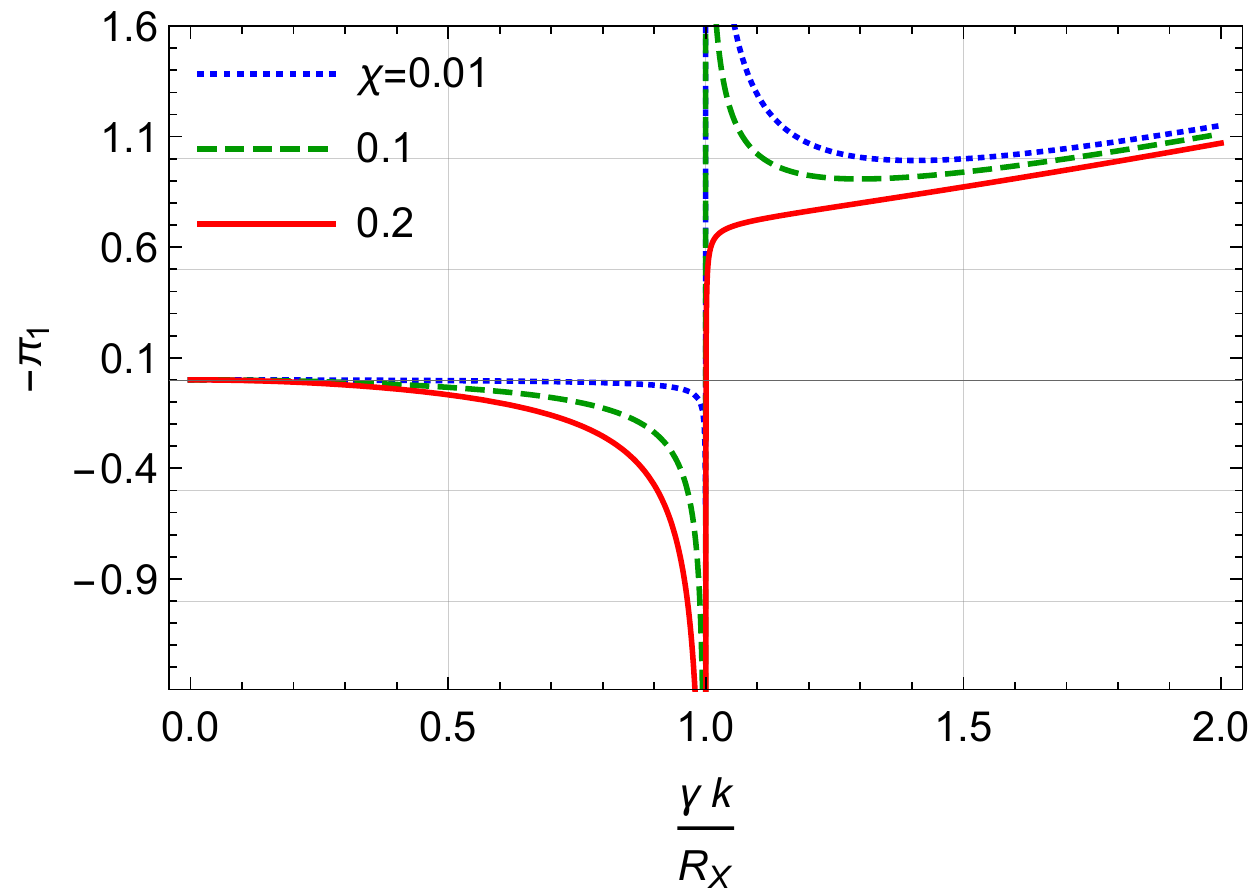}
\caption{Imaginary (upper row) and real parts (lower row) of the polarization function for a two-dimensional system with linear carrier dispersion \textit{vs.} $\gamma k/R_X$ for a quantum degeneracy of $\chi=0.05$ and several $\hbar\omega$ (left panels) and for $\hbar\omega=1\,R_X$ and several $\chi$.
}
\label{fig:repiimpilindisp}
\end{center}
\end{figure}

Imaginary and real parts of the polarization function are depicted in figure \ref{fig:repiimpilindisp} in the form
\begin{equation}\label{pi1pi2}
\Pi(k,\omega)=\frac{1}{4\beta\gamma^2}\left[\pi_1(k,\omega)+\mathrm{i}\,\pi_2(k,\omega)\right]
\end{equation}
for a temperature of $1/\beta=0.5\,R_X$ and different frequencies $\omega$ and degrees of quantum degeneracy $\chi=n\Lambda^2/2$. Note that, in this representation, $\pi_1$ and $\pi_2$ are independent on the specific material, its properties enter only via the excitonic Rydberg energy $R_X$ and the overall prefactors.

The quantities exhibit the well-known principal behavior of $\Pi$ for graphene-like systems (see, e.g., \cite{shung1986,HD07,kotov2012,wunsch2006}), see left column in figure \ref{fig:repiimpilindisp}. The actual form of the curves, however, varies sensitively with the degree of quantum degeneracy.
In the limiting case of vanishing degeneracy (equivalent to vanishing occupation of the upper cone), the remaining contributions (only from interband transitions) in (\ref{ImPilinfinal}) and (\ref{RePilinfinal}) agree with the $\mu=0$ ground state case derived, e.g., in \cite{shung1986,wunsch2006}. The corrections proportional to the degree of degeneracy modify the spectral structures both inside ($|\hbar\omega|>\gamma k$) and outside the cones ($|\hbar\omega|<\gamma k$) considerably as shown in the right column of figure \ref{fig:repiimpilindisp}.

\end{document}